\documentclass[a4paper,10pt]{article}
\usepackage[utf8]{inputenc}
\usepackage{hyperref}
\usepackage{url}
\pdfoutput=1
\usepackage{amssymb}
\usepackage{amsmath}
\usepackage{wasysym}
\usepackage{verbatim}
\usepackage[normalem]{ulem}
\newcommand{\opluslhrim}{\mathbin{\rlap{$\Leftcircle$}{+}}}

\usepackage{appendix}
\usepackage{graphicx}
\usepackage{mathrsfs}
\usepackage{caption}
\usepackage{float}
\usepackage{subfig}

\newcommand{\e}{\epsilon}

\newcommand{\p}{\partial}

\newcommand{\non}{\nonumber}

\newcommand{\be}[1]{ \begin{equation}\label{#1} }
\newcommand{\ee}{\end{equation}}
\newcommand{\bea}[1]{\begin{eqnarray}\label{#1} }
\newcommand{\eea}{\end{eqnarray}}
\newcommand{\bes}{\begin{subequations}}
\newcommand{\ees}{\end{subequations}}

\newcommand{\<}{\langle}
\renewcommand{\>}{\rangle}

\newcommand{\D}{\Delta}

\newcommand{\lb}{\Big[}
\newcommand{\rb}{\Big]}

{\title{Interacting Conformal Carrollian Theories: Cues from Electrodynamics}}
{\author{$^{[a]}${Kinjal Banerjee}$^1$, $^{[a,b]}${Rudranil Basu}$^2$,$^{[c]}${Aditya Mehra}$^3$,\\ $^{[a]}${Akhila 
Mohan}$^4$, and $^{[a]}${Aditya Sharma}$^5$}}

\date{}
\begin{document}
\maketitle
\noindent
1:\url{kinjalb@gmail.com}, 2:\url{rbasu@g.harvard.edu},
3:\url{aditya.mehra@aei.mpg.de},\\
4:\url{p20130017@goa.bits-pilani.ac.in},5:\url{p20170442@goa.bits-pilani.ac.in}\\

\noindent
[a] {{BITS-Pilani, KK Birla Goa Campus, NH 17B, Bypass Road, Zuarinagar, Goa, India 403726\\}}
[b] {Center for the Fundamental Laws of Nature, Harvard University, Cambridge, MA 02138, USA\\}
[c] {Max-Planck-Institut f{\"u}r Gravitationsphysik, Albert-Einstein-Institut, 14496, Golm, Germany\\}

\begin{abstract}
 {We construct the free Lagrangian of the magnetic sector of Carrollian electrodynamics. The 
construction relies on Helmholtz integrability condition for differential equations in a self consistent 
algorithm, working hand in hand with imposing invariance under infinite dimensional Conformal Carroll 
algebra. It requires inclusion of new fields in the dynamics and the system is free of gauge redundancies.
We next add interaction (quartic) terms to the free Lagrangian, strictly constrained by conformal 
invariance and Carrollian symmetry. The dynamical realization of the non-semi simple infinite dimensional
symmetry algebra at the level of charge algebra is exact and free from central terms.}
\end{abstract}

\newpage
\section{Introduction}

Symmetry principles play an extremely crucial role in building models describing fundamental particles and interactions. It is almost always the case that larger the symmetry group, the better is the predictive power of the theory. Conformal symmetry is one of the most useful and powerful symmetries observed in nature. 
The power of conformal symmetry is beautifully realised in two dimensions by looking at the two copies of the infinite dimensional Virasoro algebra \cite{Belavin:1984vu}. These indeed lead to integrability along with a plethora of information, e.g. calculating arbitrary correlation functions with a minimal set of data, 
using the bootstrap program \cite{Simmons-Duffin:2016gjk}.

On the other hand, organising the space of Quantum Field Theories (QFT) by classification of conformal field theories (CFT) is an alluring yet extremely challenging umbrella program. This overall wisdom is guided by the age old Wilsonian point of view, which supposedly should dictate whether a certain QFT can flow to a CFT via a relevant  deformation. In that sense, it is important that we scan for all sensible CFTs, not only those involving Lorentzian symmetry. This program has received considerable impetus in recent times, involving Galilean and Carrollian invariant CFTs \cite{Duval:2009vt,Bagchi:2014ysa,Bergshoeff:2014uea,Duval:2014uva,Duval:2014uoa,Duval:2014lpa,Bagchi:2015qcw,Bergshoeff:2015sic,Hartong:2015xda,Bergshoeff:2015uaa,Bagchi:2016bcd,Bergshoeff:2017btm,Bagchi:2017yvj,Basu:2018dub,Bagchi:2019clu,Bagchi:2019xfx,Banerjee:2019axy,Gupta:2020dtl,Chapman:2020vtn}. Interestingly both of these sectors have infinite dimensional global symmetry groups, for $d > 2$. 

The general understanding regarding these space-time backgrounds is that they can be found starting from a Minkowski one. In particular, Carrollian physics is believed to be the ultra-relativistic $(c\rightarrow 0)$ limit of Lorentz covariant physics. Effectively, the transition from Minkowski to Carroll space-time means closing up of light cone. This is directly connected with the traditional lore \cite{Bergshoeff:2014jla} that the Carroll particles \textit{don't move}. However, field theories on these space-times have extremely interesting dynamics as we will review shortly. This line of study basically stems from kinematical symmetry structures. The ultra-relativistic limit on a  relativistic conformal symmetry algebra produces the Conformal Carrollian Algebra (CCA) \footnote{Although not the main concern of our present line of investigations, we must mention that the BMS symmetry algebra \cite{Bondi:1962px,Sachs:1962zza}, that describes asymptotic symmetries of a gravitational theory on 4 (or 3) dimensional asymptotically flat space-times are conformal Carroll isometries of a 3 (or 2) dimensional Carrollian space-time \cite{Bagchi:2010eg,Duval:2014uva,Duval:2014uoa,Duval:2014lpa,Bergshoeff:2017btm}. In last 5-6 years, the BMS group has been found as a symmetry of quantum gravity S-matrix and is being related to the Weinberg's soft graviton theorem as a result of Ward identity corresponding to the symmetry \cite{Campiglia:2015yka,Strominger:2017zoo,Laddha:2017ygw,Banerjee:2018fgd,Banerjee:2019tam,Laddha:2019yaj}. }. In the special case of $d=2$, as the relativistic conformal isometries form an infinite dimensional algebra, it is plausible that CCA in 2D also has infinite number of generators. Curiously and very counter-intuitively in space-time dimensions $d>2$, CCA is infinite dimensional \cite{Bagchi:2010eg,Bagchi:2012cy,Bagchi:2013hja,Bagchi:2012xr,Bagchi:2015nca}. For the case of $d > 3$, the infinite extension is given by the Abelian ideal ($\mathcal{A}$) and the CCA becomes the semi-direct sum\cite{Bagchi:2016bcd}:  $so(d+1) \mathbin{\opluslhrim} \mathcal{A}$ of the conformal algebra of $d-1$ dimensional Euclidean space and $\mathcal{A}$. In passing we must mention that although free field theories on Minkowski space do possess infinite number of global symmetry generators (which act locally only in momentum space), only a finite subgroup (Poincare or conformal) are realized as real space-time (conformal) isometry transformations. An explicit description of an infinite set of such global symmetries has been presented in the Appendix \ref{infu}. However it is hardly possible to extend those for interacting theories.

Since conformal Carroll isometries act locally in real space, it sheds an interesting possibility of finding interacting theories with infinite dimensional symmetry group. This led some of the present authors to an ambitious program of looking into Carrollian field theories. The infinite dimensional conformal Carrollian symmetries were seen in various ultra-relativistic CFTs at the level of equations of motion in $d=4$  \cite{Bagchi:2016bcd,Bagchi:2019xfx}. These symmetry generators act locally on fields. This can be contrasted \cite{Beisert:2017pnr,Beisert:2018zxs} with the infinite hierarchy of classical Yangian symmetry generators which act non-locally on field in position space, that responsible for integrability of certain supersymmetric QFTs. 

There are two routes one can take to understand a scale invariant Carrollian field theory. One is by taking an ultra-relativistic limit of well 
understood relativistic field theories either at the level of the action or at the level of equations of motion (as we'll see later in the paper these two 
limits are not commensurate always). The second one is by building up from scratch an invariant action made of fields that have well-defined Carroll 
transformation properties. In \cite{Bagchi:2016bcd,Bagchi:2019xfx}, the authors took the first route and proposed a host of ultra-relativistic 
non-Abelian gauge theories without and with various possible matter couplings, which in $d=4$ possess infinite conformal Carrollian symmetries at
the level of equations of motion \footnote{It is to be noted, when viewed as an ultra-relativistic limit of relativistic
physics, vector fields of Minkowski space are mapped to two distinct class of fields, depending upon causality. These
classes fall into distinct representations of CCA. For historical reasons, they are called the Electric and Magnetic
sectors.}. All those systems of equations do not necessarily descend from an action. However, to have a better understanding of the classical 
dynamics and to undertake a quantization program, an action formalism for a field theory is needed.  An action of the electric sector of 
Carrollian electrodynamics \cite{Basu:2018dub} was proposed as a first example of a Carrollian field theory action. But true quantum effects in 
QFTs are only realized at 1-loop level of interacting theories. Towards this, the action formulation of Carrollian scalar electrodynamics 
(again in the electric sector) was described in \cite{Bagchi:2019clu}. 

In this paper, we take a hybrid of above two approaches, for the magnetic sector of electrodynamics in the ultra-relativistic limit whose 
equations of motion are not derivable from an action.  In this paper, we improve the situation by introducing newer fields keeping in mind the symmetry principle. The new theory, with a certain choice of newer set of marginal deformations, can be understood to be derived from action. However, in the limiting procedure, an essential feature of electrodynamics, the $U(1)$ gauge invariance is broken. The absence of gauge redundancy however is a useful feature as far as quantization is concerned. The most important feature of this example is that, it is an interacting theory invariant under the infinite dimensional conformal Carroll isometry group. And curiously, this does not descend either at the level of action, or at the level of equations of motion, 
as an ultra-relativistic limit of a relativistic field theory.
\newpage
\subsection*{Outline of the paper}
A brief summary of the paper is given below. As previously mentioned, this paper deals with finding the action for the magnetic limit of Carrollian electrodynamics. To set the stage, in section 2 we briefly review the infinite dimensional conformal Carrollian algebra and how the generators of the algebra acts on fields. We also show how that a finite number of generators constrain the correlation functions based on a set of very generic symmetry prescriptions. In section 3, we describe what precisely define Carrollian or ultra-relativistic limits of electrodynamics and difficulties encountered while writing an action for the magnetic sector of the system. Then we propose an algorithm for finding an action by introducing minimally more degrees of freedom and which do not break the conformal Carroll symmetry. In section 4, we look for a Minkowski ascendant of this new theory with new degrees of freedom where we answer the question: does this theory that we constructed come from a Lorentz invariant theory  at an ultra-relativistic limit? In section 5 we enhance the Lagrangian with the addition of interaction terms which are invariant under Carrollian symmetries and then comment on observation that the Carrollian symmetries are dynamically realized on the infinite dimensional vector space of conserved Noether charges. We end with conclusions and a discussion on the directions of future work in section 6. Our paper also has two appendices. The first one deals with the existence of infinite dimensional symmetries in relativistic theories, a fact which is not often appreciated. In the second appendix we calculate the photon propagator of electrodynamics in position space to demonstrate an interesting similarity between those and the two point correlation functions of our theory.
\newpage
%
%
%
\section{Conformal Carrollian isometries and kinematics}
Our primary goal is to construct an interacting field theory on Carroll space-time, starting from an ultra-relativistic limit of electrodynamics. As is true for any field theory, symmetry principles play a major role to constrain the kinetic as well as the interaction terms. That Carroll space-times have an infinite dimensional isometry group, as opposed to Riemann ones, is a well understood result, both from an intrinsically geometric approach \cite{Duval:2014uva} and a physically motivated one, viewing Carroll space-time as an ultra-relativistic limit of Minkowski space \cite{Bagchi:2016bcd}. In the next section, \ref{car_isometry}, we present the results for notational consistency continuity and for the sake of completeness. In the subsequent section, we provide transformation rules of fields under these isometries also developed in \cite{Bagchi:2016bcd}, now motivated from a physically intuitive point of view.
\subsection{The isometry algebra} \label{car_isometry}
One of the most straightforward ways to understand the physics and geometry of (flat) Carroll manifolds is by considering the ultra-relativistic limit of those on Minkowski space:
\be{}\label{stc}
x_i\to x_i,~ t \to \e t,~ \e \to 0.
\ee
When applied to the (conformal) isometries of Minkowski space, this amounts to an Inonu-Wigner contraction of the Poincare algebra (conformal symmetry algebra). In the Table [\ref{tab:cvb}] we summarise the generator vector fields found from the ultra-relativistic contraction of the relativistic conformal isometry generators. 
\begin{table}[H]
\begin{center}
\begin{tabular}{ |p{.5cm}|p{3cm}|p{5cm}|}
 \hline
& Transformations & Generators\\
 \hline
1.  &Translation:&$H=\p_t, ~ P_i=\p_i$\\
2.&  Rotation:& $J_{ij}=(x_i \p_j-x_j \p_i)$\\
3. &Boost:&$B_i=x_i \p_t$\\
4. &Dilatation:&$D= (t \p_t+x_i \p_i)$\\
5.& Spatial SCT:&$K_j=2x_j(t\p_t+x_i\p_i)-(x_i x_i)\p_j$\\
6.&Temporal SCT: &$K= x_i x_i \p_t$\\
 \hline
\end{tabular}
\caption{Conformal Carrollian generators. SCT stands for special conformal transformation}\label{tab:cvb}
\end{center}
\end{table}
For a $d$ dimensional space-time, the Lie algebra formed by these generators is $\mathfrak{iso}(d,1)$, 
basically reflecting the fact that it is an Inonu-Wigner contraction of the isometry generating algebra 
of AdS$_{d+1}$. This will be referred to as the finite conformal Carrollian algebra (CCA). It's a crucial 
observation \cite{Bagchi:2016bcd} that if we append this set of conformal Carroll isometry generators with
an infinite dimensional Abelian (under Lie bracket) one, the new set of generators still closes under Lie
bracket. Keeping the analogy of BMS algebra, these generators will be referred to as supertranslation (ST):
\be{}\label{mfs}
M_f=f(x^1, x^2, \dots ,x^{d-1})\p_t =: f(x) \p_t,
\ee
here $f(x)$ are arbitrary tensors transforming in irreducible representations of $\mathfrak{so}(d-1)$. The
following special cases are already given in the Table [\ref{tab:cvb}]:
\begin{equation} \label{M_special}
f(x)=
   \begin{Bmatrix} 
  1:\Rightarrow & M_f=H &  \\
   x_i :\Rightarrow & M_f=B_i &  \\
   x^2 :\Rightarrow & M_f=K. \\
   \end{Bmatrix} 
   \in \mathrm{finite~CCA}.
\end{equation}
Therefore, the infinite dimensional CCA consists of finite generators given in the Table [\ref{tab:cvb}] along with $M_f$ for arbitrary $f$.  
The Lie brackets involving the finite set $(J_{ij} , Pi, D, K_i)$ and the infinite set $M_f$ are \cite{Basu:2018dub,Bagchi:2019xfx}:
\bea{infalgebra}
&&\non [P_i, M_f] =M_{\p_i f},\quad  [D,M_f] =M_h,~\text{where}~h=x_i \p_i f-f,\\
&&\non [K_i,M_f]= M_{\tilde{h}},~\text{where}~\tilde{h}=2x_i h-x_k x_k\p_i f,\\
&&[J_{ij},M_f]= M_{\tilde{g}},~\text{where}~\tilde{g}=x_{[i}\p_{j]}f,\non\\
&&[M_f,M_{\tilde{g}}]=0.
\eea
In order to justify that the above illustrated $\mathfrak{iso}(d,1)$ with its infinite extension does indeed form the conformal isometry of a 
Carrollian manifold, we view the problem from a geometric perspective. The minimal geometrical data that specifies a Carrollian manifold $(M,g,X)$ is a rank two symmetric covariant tensor $g$ and a vector field $X$, such that $X^{\mu} g_{\mu \nu} =0 $ everywhere on the manifold. This makes the tensor $g_{\mu \nu}$ degenerate and sets apart Carroll ones from (pseudo) Riemannian ones. Conformal isometries of $(M,g,X)$ are defined as diffeomorphisms generated by vector fields $Y$, such that:
\begin{eqnarray}\label{lie}
\mathcal{L}_{Y} g= \lambda g, ~~ \mathcal{L}_{Y} X= - \frac{\lambda}{2}  X.
\end{eqnarray}
We will choose the coordinates $(t, x^1, \dots , x^{d-1} )$ for a flat Carroll manifold (which descends from Minkowski space as an ultra-relativistic limit described above), such that:
\begin{eqnarray}
g = \delta_{ij} dx^i \otimes dx^j, ~~ X = \partial_t \nonumber .
\end{eqnarray}
Solving \eqref{lie} exactly gives the infinite number of linearly independent solutions as presented in Table [\ref{tab:cvb}] and \eqref{infalgebra}. 
Henceforth this infinite dimensional algebra will be called the conformal Carrollian algebra (CCA).

\subsection{Transformation of fields under conformal Carroll isometries} \label{transfrom_car}
As we aim to study field theories with conformal Carroll symmetries, we need to understand how the above mentioned generators act on the fields. 
The physical motivations behind the choice of possible representations of the algebra are as follows.

Since all the microscopic Carrollian theories (as opposed to fluid descriptions \cite{Ciambelli:2018xat}) studied till now, are ultra-relativistic descendants of relativistic ones, it is a fair assumption that these have definite scaling \footnote{Note that, unlike Lifschitz or Schr\"{o}dinger systems, there is no scaling violation involved in going to the ultra-relativistic limit} and $SO(d-1)$ spatial rotational invariance properties. If we confine ourselves only with the scalar $\phi$ and vectors $\phi_i$ under the $SO(d-1)$ rotations, then the above considerations lead us to:
\bea{}
\mbox{Rotation: }&&\delta_{J} \phi(t,x)=\omega^{ij} (x_i \p_j-x_j \p_i ) \phi(t,x),\non  \\
&&\delta_{J} \phi_l(t,x)=\omega^{ij} [(x_i \p_j-x_j \p_i ) \phi_l(t,x)+\delta_{l[i}\phi_{j]}].\non \\
\mbox{Scaling: }&&\delta_{D} \phi(t,x)= (\D+t\p_t+x_i \p_i) \phi(t,x), \non \\&&
\delta_{D} \phi_l(t,x) = (\D+t\p_t+x_i \p_i) \phi_l(t,x).
\eea
We supplement these conditions, along with the usual space-time translation properties $\delta_H \phi = \p_t \phi, \delta_{P} \phi = a^i\p_i \phi$ for a constant vector $a^i$. However, this does not fix completely the action of all other generators on these fields of definite scale and spin. Motivated from the fact that we are interested in  ultra-relativistic limits of Lorentzian theories of $SO(d-1,1)$ vectors, we confine ourselves to only $SO(d-1)$ scalars and vectors to describe Carrollian ones. This essentially boils down to considering only those transformations which preserves the module of these fields and their derivatives. 

Now, let us look at the role of boost $B_i$ on fields $(\phi,\phi_l)$. Towards this, we consider the ultra-relativistic limit of the $SO(d-1,1)$ Lorentz boost transformation on a Lorentz covariant $d$ vector $ \phi_{\mu}$:
\bea{}\label{gdh}\delta_{L}\phi_{\rho}(x)=\omega^{\mu\nu}[(x_{\mu}\p_{\nu}-x_{\nu}\p_{\mu})\phi_{\rho}(x)+\eta_{\rho[\mu}\delta^{\sigma}_{\nu]}\,\phi_{\sigma}(x)],
\eea
where $\omega^{\mu\nu}$ is parameter for Lorentz transformation. 

The ultra-relativistic limit of this transformation rule works \cite{Bagchi:2016bcd} by taking appropriate limits on $\phi_{\mu}$ alongside \eqref{stc} and breaking Lorentz invariance by inhomogeneous scaling rules. However similar the case of arriving at Galilean transformation rules \cite{Levy1965}, here also one encounters a couple of possibilities depending on whether $\phi_{\mu}$ is space-like or time-like:
\bes \label{lims}
\bea{}
 \quad \phi \to \phi, \, \phi_i \to \e \phi_i \label{elimcont},\\
 \quad \phi \to \e \phi, \, \phi_i \to \phi_i \label{mlimcont}
\eea
\ees
conventionally noted respectively as the electric type and the magnetic type limits. 

Now let us consider only the boost part of the Lorentz transformation \eqref{gdh}, ie take $\omega_{0i} = b_i, \omega_{ij} = 0$ for some constant $SO(d-1)$ vector $b$. Then employing both the ultra-relativistic limits on the space-time coordinates \eqref{stc} and on the fields \eqref{elimcont} and \eqref{mlimcont} we get respectively the Carrollian boost transformation conditions:
\bes{}\label{bst}
\bea{}&&\hspace{-.5cm}\delta_{B} \phi(t,x)=b^j[x_j\p_t \phi(t,x)],~\delta_{B}\phi_l(t,x)=b^j[x_j\p_t \phi_l(t,x)- \delta_{lj} \phi(t,x)],\label{bst1}\\&&\hspace{-0.5cm}
\delta_{B} \phi(t,x)=b^j[x_j\p_t \phi(t,x)-\phi_j(t,x)],~\delta_{B}\phi_l(t,x)=b^j[x_j\p_t \phi_l(t,x)].\label{bst2}
 \eea\ees
A more compact notation for the above equation \eqref{bst} is
\bea{}\label{defqq}&&
\delta_{B} \phi(t,x)=b^j[x_j\p_t \phi(t,x)+q\phi_j(t,x)],\non\\&&\delta_{B}\phi_l(t,x)=b^j[x_j\p_t \phi_l(t,x)+q^\prime \delta_{lj} \phi(t,x)].
\eea
We get \eqref{bst1}, if we take the value of the constants as $(q=0,q'=-1)$ and for $(q=-1,q'=0)$, we get \eqref{bst2} which are respectively be referred to as \textit{the electric and the magnetic sector}. 
It can be easily checked that these are exactly the same as the electric and magnetic sector defined in 
(\ref{lims}).

Similarly, the action of $(K,K_i)$ can be found by taking the Carrollian limit on relativistic special conformal transformation. The final results become
\bes{}
\bea{}
&&\hspace{-1.3cm}\delta_{k}\phi(t,x)=k[x^2\p_t\phi(t,x) +2qx_i\phi_i(t,x)],~\delta_{k}\phi_l(t,x)=k[x^2\p_t\phi_l(t,x) \nonumber\\
&& \quad +2q^{'}x_l\phi(t,x)],\\
&&\hspace{-1.3cm}\delta_{k} \phi(t,x)= k^j \lb(2\D x_j+2x_jt\partial_t+2x_i x_j \partial_i- x_i x_i \partial_j )\,\phi(t,x)+2t q \phi_j(t,x)\rb,\\
&&\hspace{-1.3cm}\delta_{k} \phi_l(t,x)= k^j \lb(2\D x_j+2x_jt\partial_t+2x_i x_j \partial_i- x_i x_i \partial_j )\rb\,\phi_l(t,x)\non\\
&&\hspace{2cm}+2k_l x_j \phi_j(t,x)-2k_i x_l \phi_i(t,x)+2tq^\prime k_l \phi(t,x).
\eea
\ees
We are only left with the action of infinite number of generator $M_f$ \eqref{mfs} on the fields. For supertranslations, we don't have any relativistic counterpart. But an ansatz for these transformations can be given, motivated by the transformations of the fields under $H,B_i,K$, as these generators are special cases of $M_f$ \eqref{M_special}. We first  conjecturally state the transformation rules:
\bea{} \label{M_general}
&&\delta_{M_f} \phi(t,x)= f(x)\p_t \phi(t,x)+q \phi_i(t,x)\p_i f(x),\\
 &&\delta_{M_f} \phi_l(t,x)= f(x)\p_t \phi_l(t,x)+q^\prime \phi(t,x)\p_l f(x).
\eea
That the above conjecture is consistent, can be verified as follows. Let $A,B \in \mathfrak{g}$ be a symmetry algebra, which in our case is the conformal Carrollian one and let $\delta_A$ etc. denote infinitesimal change on fields ie. transformations as given in \eqref{M_general}. Then if the commutator relation (as the difference of alternated consecutive transformations):
\bea{}
[ \delta_A , \delta_B ] = \delta_{[A,B]}
\eea
on the space of fields of interest holds, we say the transformation rules are self-consistent\footnote{At a more formal level this is a statement of homomorphism from the lie algebra $\mathfrak{g}$ to that of the algebra of vector fields on field space.}.

It can be easily checked that with the above ansatz of transformation \eqref{M_general}, the self consistency of conformal Carroll transformation holds, 
thus validating the ansatz. We also note that this ansatz of course is not the most general one. However, for the purpose of the present paper, 
ie. to see Carrollian transformation of $SO(d-1)$ scalars and vectors, this is sufficient. 
For ease of reading, we have collated all the transformations in Table [\ref{tab:cvc}].
\begin{table}[H]
\begin{center}
\begin{tabular}{|p{2cm}|p{11cm}|}
 \hline
Translation:&$\delta_{p} \phi(t,x)=p^j \p_j \phi(t,x)$\\
       &$ \delta_{p} \phi_l(t,x)=p^j \p_j \phi_l(t,x)$\\ 
     \hline   
Rotation:& $\delta_{\omega} \phi(t,x)=\omega^{ij} (x_i \p_j-x_j \p_i ) \phi(t,x)$\\
 &$ \delta_{\omega} \phi_l(t,x)=\omega^{ij} [(x_i \p_j-x_j \p_i ) \phi_l(t,x)+\delta_{l[i}\phi_{j]}]$\\
 \hline
Boost:&$\delta_{B} \phi(t,x)=b^j[x_j\p_t \phi(t,x)+q\phi_j(t,x)]$\\
& $\delta_{B}\phi_l(t,x)=b^j[x_j\p_t \phi_l(t,x)+q^\prime \delta_{lj} \phi(t,x)]$\\
\hline
Dilatation:&$\delta_{\Delta} \phi(t,x)= (\D+t\p_t+x_i \p_i) \phi(t,x)$\\
&$\delta_{\Delta} \phi_l(t,x)= (\D+t\p_t+x_i \p_i) \phi_l(t,x)$\\
\hline
SCT:&$\delta_{k} \phi(t,x)= k^j \lb(2\D x_j+2x_jt\partial_t+2x_i x_j \partial_i- x_i x_i \partial_j )\,\phi(t,x)+2t q \phi_j(t,x)\rb$\\
&$\delta_{k} \phi_l(t,x)= k^j \lb(2\D x_j+2x_jt\partial_t+2x_i x_j \partial_i- x_i x_i \partial_j )\rb\,\phi_l(t,x)$\\
&$\hspace{4cm}+2k_l x_j \phi_j(t,x)-2k_i x_l \phi_i(t,x)+2tq^\prime k_l \phi(t,x)$\\
\hline
ST:&$\delta_{M_f} \phi(t,x)= f(x)\p_t \phi(t,x)+q \phi_i(t,x)\p_i f(x)$\\
&$ \delta_{M_f} \phi_l(t,x)= f(x)\p_t \phi_l(t,x)+q^\prime \phi(t,x)\p_l f(x).$\\
\hline
\end{tabular}
\caption{Transformation of fields under CCA}\label{tab:cvc}
\end{center}
\end{table}
Now that we have understood the conformal Carrollian symmetry generators, with the power of conformal symmetry we can constrain two point functions of 
fields which transform according to Table [\ref{tab:cvc}] in the next section. For that we require a definition of a unique vacuum state which respects 
the global symmetries: spatial and temporal translations, Carrollian boost, spatial rotations, dilatation and Carrollian special conformal transformation.

\subsection{Correlation Functions}\label{cf} 
Conformal symmetry is a powerful tool, because this helps us find 2 or 3 point functions of a set of fields with definite conformal transformation 
properties irrespective of the existence of a Lagrangian description. Here we would explore this idea for fields with above mentioned conformal Carroll 
transformations. We will make a very generic and plausible assumption of the existence of a vacuum state, that is invariant under the global part of 
conformal Carroll algebra.
 
To set up the context,
\begin{itemize}
\item[1.] Let $ \Phi(t,x)$ and $ \tilde{\Phi}(t,x)$ be fields which transform either as scalar or vector under $SO(d-1)$ and has definite scaling dimensions. 
\item[2.] Let's assume any global generator $ \bigstar$ being a symmetry of the vacuum, would mean $ \bigstar |0 \> =0$ and $ \< 0 | \bigstar =0  $.
\item[3.] If we use it in the context of correlators, it gives:
$$ \< 0 |[ \bigstar , \Phi (t_1,x_1) ] \, \tilde{ \Phi }(t_2, x_2) | 0 \> + \< 0 |  \Phi (t_1,x_1) \, [ \bigstar , \tilde{ \Phi }(t_2, x_2) ] | 0 \>=0 $$  
\end{itemize}
This will give a set of simultaneous differential equations of correlation functions. The solutions to these equations will give us the required correlation functions. 
\subsubsection*{Case 1: When both $\phi$ and $\phi_i$ are in electric sector}
\begin{itemize}
\item  $G_{00}(t,x)\equiv\langle \phi (t_1, x_1) \,\phi (t_2, x_2) \rangle:$\\
We first consider both $\phi(t,x)$ in electric sector, ie. $(q=0, q'=-1)$ as per the notation of \eqref{defqq}. We will now impose the invariance under $H,P_i,J_{ij},D$ transformation. The results which follows from the differential equations is
\bea{}
G_{00}(t,x) = \sum _{m \in \mathbb{Z}}\alpha_{m} t^{m} r^{-m-2},
\eea
where $x^i = x^i_1 - x^i_2, t= t_1 - t_2$ and $r^2 = x^ix_i$ and $n=-(m+2)$. 
Now, let's impose the invariance under $B_{i}$:
\bea{} 
\sum _m m\, \alpha_m \, t^{m-1} r^{-m-2} x_i=0,
\eea
The solution to this equation gives $\alpha_{m}=0 \ \forall \ m \neq 0$. Hence, using the constraint provided by $B_{i}$ gives us the correlation function:
\bea{} G_{00}(t,x)\equiv\langle \phi (t_1, x_1) \phi (t_2, x_2) \rangle =  \frac{\alpha}{r^2}.\eea
As expected, the Invariance under $(K_i, K)$ give nothing new and simply respect this form of the correlation function. 
\item $ G_{0i}(t,x)\equiv \langle \phi (t_1, x_1) \phi_i (t_2, x_2)\rangle :$\\
Applying the above scheme, we again impose the invariance of the vacuum under $H, P_i,J_{ij},D$. The result dictates: 
\bea{} G_{0i}(t,x) =  \sum _{m}\beta_{m} \, x_i \, t^{m} r^{-m-3},\eea
where the above expression comes only when we take $m+n+3=0$.
Let's now implement the invariance under $B_{i}$, we get
\bea{}
x_l\p_t G_{0i}-\delta_{li}G_{00} =0   \Rightarrow \beta_m =0, ~ \forall \ m \ \text{and} \ \alpha=0.
\eea
Both $G_{00}$ and $G_{0i}$ vanish completely. 
\item $  G_{ij}(t,x)\equiv\langle \phi_i (t_1, x_1)\phi_j (t_2, x_2) \rangle:$\\
The expression for the correlation function after we impose the invariance under $H, P_i, J_{ij}, D$ becomes:
\bea{}
G_{ij}(t,x) =  \sum _{m}t^m r^{-m-2}\Big[\gamma^1_{m}\delta_{ij}+\gamma^2_m \frac{x_ix_j}{r^2}\Big].
\eea
Imposing the invariance under $B_l$, we get the constraint as
\bea{}
\sum_{m}t^{m-1}r^{-m-2}m x_l \Big[\gamma^1_{m}\delta_{ij}+\gamma^2_m \frac{x_ix_j}{r^2}\Big]=0 \Rightarrow (\gamma^1_m, \gamma^2_m)=0 ~\forall ~m \neq 0.
\eea
Therefore, the final result become
\bea{}
G_{ij}(t,x)\equiv \langle \phi_i (t_1, x_1)\phi_j (t_2, x_2) \rangle  = \frac{\gamma_1}{r^2} \delta _{ij}+\frac{\gamma_2}{r^4}x_ix_j.
\eea
Imposing the invariance under $(K_i, K)$ on the correlation function gives nothing new.
\end{itemize}
\subsubsection*{Case 2: When both $\phi$ and $\phi_i$ are in magnetic sector}
\begin{itemize}
\item $K_{ij}\equiv \langle  \phi_i (t_1, x_1)\phi_j (t_2, x_2) \rangle:$\\
Imposing invariance under $H, P_i, J_{ij}, D$ and $B_i$, restricts the correlation function to:
\bea{} 
K_{ij}\equiv \langle  \phi_i (t_1, x_1)\phi_j (t_2, x_2) \rangle =  \frac{\rho_1}{r^2} \delta _{ij}+\frac{\rho_2}{r^4}x_ix_j.
\eea 
The form of the correlation function remains same even if we impose $K,K_i$.
\item $K_{0i}\equiv \langle  \phi (t_1, x_1)\phi_i (t_2, x_2) \rangle:$\\
Looking for invariance under $H,P_i,J_{ij},D$, the correlation function becomes
\bea{} K_{0i}(t,x) =  \sum _{m}\sigma_{m} \, x_i \, t^{m} r^{-m-3},\eea
After the implementation of invariance under $B_l$, we get 
\bea{}
K_{0i}(t,x)=0,~K_{ij}(t,x)=0.
\eea
\item $K_{00}\equiv \langle  \phi (t_1, x_1)\phi (t_2, x_2) \rangle:$\\
The correlation function $K_{00}$ after we impose the invariance under \\
$H,P_i,J_{ij},D,K,K_i$ becomes
\bea{}K_{00}\equiv \langle  \phi (t_1, x_1)\phi (t_2, x_2) \rangle=\frac{\sigma}{r^2}\eea
\end{itemize}
\subsubsection*{Case 3: When $\phi$ and $\phi_i$ are in either electric or magnetic sector}
\begin{itemize}

\item $H_{0i}\equiv\langle \phi(t_1, x_1)\phi_i (t_2, x_2) \rangle:$
\\
For this case, we take $\phi$ in electric sector and $\phi_i$ in magnetic sector. We will impose the invariance under $H, P_i, J_{ij}$ and $D$ to get the form of correlator as:
\bea{}
H_{0i} =  \sum _{m \in \mathbb{Z}} e_m t^{m} \, r^{-m-3} \, x_i. 
\eea
The constraint equation which we get after we impose the invariance under $B_l$, is given by
\bea{}\sum _{m \in \mathbb{Z}} e_m \, m t^{m-1} r^{-m-3}\, x_i x_{l}=0 \Rightarrow e_{m}=0 \ \forall \ m \neq 0.
\eea
The final expression of the correlation function becomes
\bea{}
H_{0i}\equiv\langle \phi(t_1, x_1)\phi_i (t_2, x_2) \rangle= \frac{e}{r^{3}} x_i. 
\eea
This form remains unchanged under the invariance of $K_l,K$.
\item $H_{ij}\equiv \langle \phi_i (t_1, x_1)\phi_j (t_2, x_2) \rangle:$\\
Here, we take one of $\phi_i$ in electric sector and another $\phi_j$ in magnetic sector. Following the same procedure, we will first impose the invariance under $H, P_i, J_{ij}$ and $D$. Next, we will
\bea{}
H_{ij} = \sum _{m \in \mathbb{Z}} f_m \,  t^{m} r^{-m-2}\, \delta_{ij}.
\eea
now impose the invariance under $K$, we get
\bea{}
x^2\p_tH_{ij}-2x_j H_{i0}=0.
\eea
which implies that $H_{ij}$ and $H_{0i}$ vanishes completely.
\item  $H_{00}(t,x)\equiv\langle \phi (t_1, x_1) \,\phi (t_2, x_2) \rangle:$\\
We take first $\phi$ in electric sector and latter $\phi$ in magnetic sector.We will now impose the invariance under $H,P_i,J_{ij},D$ transformation. The correlation function becomes
\bea{}
H_{00}(t,x) = \sum _{m \in \mathbb{Z}}\theta_{m} t^{m} r^{-m-2},
\eea
Using the constraint provided by $K$ gives us the correlator:
\bea{} H_{00}(t,x)\equiv\langle \phi (t_1, x_1) \phi (t_2, x_2) \rangle =  \frac{\theta}{r^2}.\eea
\end{itemize}
The summary of the correlation functions are given in Table [\ref{tab:2pt}].
\begin{table}[H]
\begin{center}
\begin{tabular}{|p{0.9\linewidth}|}\hline 
\rule{0pt}{5ex}
\underline{Case 1: $\phi$ and $\phi_i$ are in electric sector:} \\
$  \langle \phi (t_1, x_1) \phi (t_2, x_2) \rangle=0,~\langle \phi (t_1, x_1)\phi_i (t_2, x_2)\rangle=0,  $\\
$\langle \phi_i (t_1, x_1)\phi_j (t_2, x_2) \rangle=\frac{\gamma_1}{r^2} \delta _{ij}+\frac{\gamma_2}{r^4}x_ix_j.$\\
\underline{Case 2: $\phi$ and $\phi_i$ are in magnetic sector:}\\
$  \langle \phi (t_1, x_1) \phi (t_2, x_2) \rangle=\frac{\sigma}{r^2},~\langle \phi (t_1, x_1)\phi_i (t_2, x_2)\rangle=0,  $\\
$\langle \phi_i (t_1, x_1)\phi_j (t_2, x_2) \rangle=0.$\\
\underline{Case 3: $\phi$ and $\phi_i$ are in either electric or magnetic sector:}\\
$ \langle \phi (t_1, x_1)\phi_i (t_2, x_2) \rangle=0,~
 \langle \phi_i (t_1, x_1)\phi_j (t_2, x_2) \rangle=0.$\\
  $\langle \phi (t_1, x_1)\phi (t_2, x_2) \rangle=\frac{\theta}{r^2}.$\\
\hline
\end{tabular}
\caption{Summary of the Two Point Correlation Functions}\label{tab:2pt}
\end{center}
\end{table}
We can also write the correlation functions in most generalized manner. They are given by
\bes{}\label{sea}
\bea{}&&
\hspace{-1cm}\langle 0| \phi (t_1,x_1) \phi_l (t_2, x_2 )|0\rangle=0,\label{sea1}\\&&
\hspace{-1cm}\langle 0| \phi (t_1,x_1) \phi (t_2, x_2 )|0\rangle=\big(a_1(q_e+q'_{e})+b_1(q_m+q'_{m})+2\big)\,\frac{\tilde{\gamma}}{r^2},\\&&
\hspace{-1cm}\langle 0| \phi_i (t_1,x_1) \phi_j (t_2, x_2 )|0\rangle=\big(a_2(q_e+q'_{e})+b_2(q_m+q'_{m})+2\big) \nonumber \\
&&\hspace{3cm}\Big[\frac{\gamma}{r^2} \delta _{ij}+\frac{\gamma'}{r^4}x_ix_j\Big].
\eea\ees
In the correlators above \eqref{sea}, when the fields transform in electric sector, we take the constants as $(a_1=2,a_2=1, b_{1,2}=0)$ along with $(q_e=0,q'_e=-1)$ and for magnetic case, $(a_{1,2}=0,b_1=1,b_2=2)$ along with $(q_m=-1,q'_m=0)$. For mixed case, we have to take $b_{1}=0,(a_1,a_2,b_{2}=1)$. 
\section{Towards a Lagrangian formulation for the magnetic limit}
While trying to formulate an ultra-relativistic limit of electrodynamics, we keep in mind that fundamental dynamical variables are still $(A_t, A_i)$, as these satisfy the Bianchi identity $F = dA = 0$, which is topological (background independent) and should hold even on Carroll manifold. In the following we first describe the problems associated with the dynamics of the magnetic sector of Carrollian sector. In the later sections we propose an algorithm to cure that and find an action principle for the same.
\subsection{Brief review on Carrollian electrodynamics}
As already hinted above, there are a couple of ways of thinking about the effect of ultra-relativistic limit of a Lorentz covariant vector field. In particular, in the case of formulating electrodynamics on a Carroll manifold, there are a couple of ways to do that \cite{Bagchi:2016bcd}, as argued in \eqref{lims}:
\begin{eqnarray} 
A_{t} \rightarrow A_t &,& A_i \rightarrow \e A_i  ~ ~ ~ ~ \Rightarrow{\mbox{Electric Limit,}}\label{els}\\
A_{t} \rightarrow \e A_t &,& A_i \rightarrow  A_i ~ ~ ~ ~ ~ \Rightarrow{\mbox{Magnetic Limit.}}\label{mls}
\end{eqnarray}
The nomenclature \cite{Duval:2014uoa} of one limit being electric and the other as magnetic is inspired from Galilean electrodynamics \cite{Levy1965} and is same as the ones given in (\ref{lims}).

The dynamics for the Carrollian electric limit is straightforward and has been worked out in \cite{Duval:2014uoa} and later in more detail in \cite{Basu:2018dub}. The equations of motion for the same are given by:
\bea{}\label{ed} \p_i\p_iA_t-\p_i\p_t A_i=0,~ \p_t\p_i A_t -\p_t\p_t A_i=0\eea
which are the Euler Lagrange equations coming from the Lagrangian (in 3 spatial dimensions):
\bea{}
L = \int d^3\,x (\p_t A_i - \p_i A_t)^2.
\eea
However the case of magnetic limit is fraught with ambiguities. For example, when one takes the ultra-relativistic limit of the Maxwell's equation: $\p_{\mu} F^{\mu \nu} = 0$, one lands up to the following spatial and temporal equations:
\bea{}\label{md} \p_i\p_t A_i=0,~\p_t\p_t A_i=0. 
\eea
$A_t$ dropping out of the equations makes it completely unrestricted in the phase space of Carrollian electrodynamics. Moreover it's evident that both of the equations \eqref{md} can't come from an action via variational principle. On the other hand, naively taking the ultra-relativistic magnetic limit on Maxwell Lagrangian results into:
\bea{}
L = \int d^3\,x (\p_t A_i)^2.
\eea
This gives rise to the equation $\p^2_t A_i =0$. One of the principal goals of this article is to construct an action principle for the magnetic limit of Carrollian electrodynamics and understand the corresponding 
dynamics better. As an additional feather, one notices her gauge invariance is lost both at the level of  \eqref{md} and the Lagrangian.

Irrespective of the existence or non-existence of an action, both the sectors \eqref{ed} and \eqref{md} were checked to be invariant under the infinite dimensional conformal Carrollian symmetry algebra \cite{Bagchi:2016bcd}. In order to check the invariance of an equations of motion of the form
$f(A,\partial A, \p^2 A) = 0$ with respect to  a symmetry generator $Q$, we would require the variational derivative equation 
\bea{}\delta_Q f(A,\p A,\p^2A) = 0\eea to hold. The explicit expressions of the variational actions of the generators are given in Table [\ref{tab:cvc}].
The invariance under space-time translations and spatial rotations are straightforward. To get the invariance under dilatation $D$, one requires the value of the scaling weight $\Delta=1$. Similarly, the invariance of equations (\ref{ed})-(\ref{md}) under SCT can be seen by using the values of the constants $(q=0,q'=-1)$ for electric sector and $(q=-1,q'=0)$ for the magnetic one.
\subsection{The Helmholtz Conditions and the Consistency Algorithm}
It's not always the case that non-Lorentz invariant theories, viewed as particular limits of relativistic equations of motions, have a consistent dynamical description in terms of action formulation. Apart from the above encountered example of ultra-relativistic limit, electrodynamics on Newton-Cartan space-time (Galilean limit) also suffers from the similar fate. For the later, a particular intelligent guess of adding an extra field \cite{Bergshoeff:2015sic} in the system of equations of motion made them such that an action formulation was plausible. Later in \cite{Banerjee:2019axy}, a systematic analysis was done and an algorithm was presented which justifies the procedure of addition of new degrees of freedom.

One set of the key ingredients in this algorithm are the Helmholtz conditions for the equations of motion. If these conditions are met, then it is guaranteed that there exists an action functional of the fields, the variational extremization of which gives these equations and that the equations are Euler Lagrange equations. In mathematics literature, this inverse problem of calculus of variations has been well studied \cite{davis2:1929,jessedouglas:1941}. We recapitulate the conditions for easy reference below. 

To begin with, we will consider a theory which is described in terms of fields $u^B$ . We will then denote the equations of motion by $T_A$, where ($A,B, \dots=1,2,...N$). In order for an action functional $S[u^B] = \int d^n x \mathcal{L}(u^B,u_a^C,u^D_{ab},x^a)$ corresponding to these equations of motion to exist, the necessary and sufficient conditions are given by the Helmholtz conditions \cite{Henneaux:1984ke}
\begin{subequations}\label{helmholtz}
\begin{eqnarray}
 \frac{\partial T_A}{\partial(u^B_{ab})} &=& \frac{\partial T_B}{\partial (u^A_{ab})} \label{helmholtz1}\\
 \frac{\partial T_A}{\partial {u^B_a}} + \frac{\partial T_B}{\partial {u^A_a}} &=& 2 \partial_b \frac{\partial
T_B}{\partial ( u^A_{ba}) } \label{helmholtz2} \\
 \frac{\partial{T_A}}{\partial {u^B}} &=& \frac{\partial T_B}{\partial u^A} - \partial_a \frac{\partial T_B}{\partial
{u^A_a}} + \partial_a \partial_b {\frac{\partial T_B}{\partial (u^A_{ab})}} \label{helmholtz3}
\end{eqnarray}
\end{subequations}
where $u_a^A$and $u_{ab}^A$ denotes the first and second derivatives of $u^A$. 

It's very clear that now using \eqref{helmholtz} that \eqref{md} don't come from an action. In order to proceed further we will carry out the following steps systematically as a consistency algorithm.

\begin{enumerate}
\item  The equations will be first passed through the Helmholtz criteria. If the criteria are satisfied by the EOMs, we then go down to step \ref{gicheck}. Otherwise we go to step \ref{step2}.
\item \label{step2} We introduce new $SO(3)$ scalars and vectors of mass dimension 1, in turn to the system of equations, which transform as by the rules of table \ref{tab:cvc}. We will add terms which are second derivatives of the new fields in space-time coordinates. 
\item \label{step_later} The new sets of equations of motion, with new terms of arbitrary coefficients will be passed through the Helmholtz condition. If for any choice of coefficients, the Helmholtz conditions are satisfied, we go to the next step, or go back to \ref{step2}.
\item \label{step3} We will further constrain the set of equations thus found by requiring them to give back the Carrollian electrodynamics equations when the newly introduced field(s) is (are) set as constant non-dynamical.
\item \label{gicheck} Finally, conformal Carrollian symmetry of the equations will be checked, which will further constrain the terms in the system of equations.
\end{enumerate}
A caveat about the point \ref{step2} above is that for now we keep on adding only fields of spin $0$ and $1$ (keeping in mind that we started off as a limit of electrodynamics). However, in principle, there is no foreseeable problem in including tower of higher spins, as long as this remains a free theory.

Let us start with the EOMs of magnetic limit of Carrollian Electrodynamics denoted as
 \begin{eqnarray} \label{mageom}
  \tilde{T}_0 := \partial_j \partial_t A_j=0,~~
  \tilde{T}_i := \partial_t \partial_t A_i=0 .
 \end{eqnarray}
 \noindent
Since they obviously do not obey the Helmholtz conditions \eqref{helmholtz} they cannot appear as Euler Lagrange
equations of motion derived from any local action. We then move on to step \ref{step2} of the above procedure and add a
minimal set of additional fields $(B_i, B_t)$\footnote{Addition of a single extra
scalar or a single extra vector field does not satisfy the algorithm given above.}.

We now consider the most general set of equations of motion of the fields $A_t$ and $A_i$ with terms corresponding to
extra scalar field $B_t$ and $B_i$. The most general second order differential equations involving these fields are given by
\begin{subequations} \label{generaleom}
\bea{} &&\hspace{-.8cm}
 T_0 := a_1 \partial_j \partial_j A_t  +  a_2 \partial_t   \partial_t  A_t  +  b_1 \partial_j \partial_t A_j  +  c_1 \partial_j \partial_j B_t  +  c_2 \partial_t \partial_t B_t \nonumber \\
 && \hspace{3.6cm} +  d_1 \partial_j \partial_t B_j = 0,   \\&&\hspace{-.8cm}
 T_i := a_3 \partial_i \partial_t A_t  +  b_2 \partial_t \partial_t A_i  + b_3 \partial_j \partial_j A_i + b_4 \partial_j \partial_i A_j + c_3 \partial_i \partial_t B_t   \non \\&&\hspace{-.8cm}
\hspace{3.6cm}+ d_2 \partial_t \partial_t B_i + d_3 \partial_j \partial_j B_i + d_4 \partial_j \partial_i B_j = 0,  \\ &&\hspace{-.8cm}
 T_B := a_4 \partial_j \partial_j A_t + a_5 \partial_t \partial_t A_t + b_5 \partial_j \partial_t A_j + c_4 \partial_j \partial_j B_t + c_5 \partial_t \partial_t B_t \non \\
 && \hspace{3.6cm} + d_5 \partial_j \partial_t B_j =0,  \\&&\hspace{-.8cm}
 T_{B_i}: = a_6 \partial_i \partial_t A_t  +  b_6 \partial_t \partial_t A_i  + b_7 \partial_j \partial_j A_i + b_8 \partial_j \partial_i A_j + c_6 \partial_i \partial_t B_t  \non \\&&\hspace{-.8cm}
 \hspace{3.8cm}+ d_6\partial_t \partial_t B_i + d_7 \partial_j \partial_j B_i + d_8\partial_j \partial_i B_j = 0. 
\eea
\end{subequations}
We now crank the machine of passing these equations through the Helmholtz criteria and find the constraints on the coefficients appearing in \eqref{generaleom}. The constraints come out to be
\begin{eqnarray}\label{helmpara}
&& b_1=a_3 ,\ c_1=a_4 ,\ c_2=a_5 ,\ d_1=a_6 ,\ c_3=b_5, \nonumber \\&&
 d_2=b_6 ,\ d_3= b_7,\ d_4=b_8 ,\ c_6=d_5.
\end{eqnarray}
Next we move on to the step \ref{step3} of the algorithm above and check the conditions found by demanding that setting  
$A_t, B_t $ and $B_i$ as constant background fields in \eqref{generaleom} with the parameters constraints \eqref{helmpara} would give us back \eqref{md}. 
This gives rise to the following further constraints:
\bea{}\label{limitpara}
    \{a_1, a_2, b_1, b_2, b_3, b_4, c_1, c_2, d_1, d_3, d_4\}=0,\{c_3,d_2\}=1.
\eea
This interestingly lets us get rid of the field $A_t$ from system of fields, as its coefficient gets to vanish and effectively we have 3 dynamical equations of motion:
\begin{eqnarray}\label{dfd}&&
 T_i := \partial_i \partial_t B_t + \partial_t \partial_t B_i =0, \nonumber\\&&
 T_B := \partial_j \partial_t A_j + c_4 \partial_j \partial_j   B_t + c_5 \partial_t \partial_t B_t + d_5 \partial_j \partial_t B_j =0,  \nonumber \\&&
 T_{B_i} := \partial_t \partial_t A_i + d_5 \partial_i \partial_t B_t + d_6\partial_t \partial_t B_i + d_7 \partial_j \partial_j B_i + d_8\partial_j \partial_i B_j = 0. 
\end{eqnarray}
\noindent
There are still undetermined constants. As per the last step of the algorithm spelled out above, we need to check whether the above equations are  invariant under conformal Carrollian transformations.  For that, we use the Table [\ref{tab:cvc}] to look for the invariance of \eqref{dfd}. Finally, we end up with the set of equations which are invariant under Helmholtz conditions and CCA with restrictions on the values of the parameters $\Delta = 1$ for both $B_t, B_i$ and boost transformation rules: $q= 0 ,q' = -1$ in the context of the table  [\ref{tab:cvc}]. However the vector field we started off originally with however transform with in $q' = 0$, as evidently the $SO(3)$ scalar $A_t$ drops off from the system of equations.  The final equations of motion in the magnetic limit given by,
\begin{eqnarray}\label{rer}&&
 T_i := \partial_i \partial_t B_t + \partial_t \partial_t B_i = 0, \nonumber\\&&
 T_B := \partial_j \partial_t A_j + c_5 \partial_t \partial_t B_t = 0, \nonumber \\&&
 T_{B_i} := \partial_t \partial_t A_i = 0 .
 \end{eqnarray}
Note that $c_5$ is an undetermined parameter which can take any arbitrary value.

As a summary, note that we have arrived at a system of equations, which give the ultra-relativistic limits of Maxwell's equations in the magnetic limit. Moreover, these equations are invariant under the infinite dimensional conformal Carrollian algebra and can be derived by variational principle from an action. Before going on to write the action, let's discuss aspects of the non-triviality of the infinite dimensional global symmetry aspect in the following.
\subsection{Strong invariance check and Lagrangian}
The following short exercise, would better illuminate the meaning of the symmetries of the equations of motion.

It is evident in a class of systems, there are transformations which are symmetries of the equations of motion, but not of the action. As described in \cite{Beisert:2018zxs}, these are characterised as weak symmetries as opposed to standard Noetherian, ie. strong ones. Weak invariance does not necessarily lead to conserved quantities are rather non-dynamical conditions on equations of motion.
 
Let us denote equations of motion, derivable from an action functional $S[\Phi^I, \p \Phi^I]$ as:
\bea{} \label{eomq} T_I:= \frac{\delta S}{\delta \Phi^I} =0. \eea
If $\star$ is generic continuous symmetry generator, ie. the symmetry condition (which should hold true off-shell) can be expressed as :
\bea{}\label{sinv} \delta_{\star} S=\int d^{d}x~(\delta_{\star} \Phi^I) T_I=0. \eea
Let's take another variational derivative of \eqref{sinv}, now with respect to $\Phi^K (y)$, to have:
\bea{}
\int d^{d}x~\left[\frac{\delta (\delta_{\star} \Phi^{I}(x))}{\delta \Phi^{K}(y)}~T_I(x) + \delta_{\star} \Phi^{I}(x) \frac{\delta T_I (x)}{\delta \Phi^K (y)} \right]=0. \nonumber
\eea
The second term in the above is $\int d^d x~ \delta_{\star} \Phi^{I}(x) \dfrac{\delta^2 S}{\delta \Phi^I (x) \delta \Phi^K(y)} = \delta_{\star} \dfrac{\delta S}{\delta \Phi^K (y)} = \delta_{\star} T_K (y)$. Hence it trivially follows that:
\bea{}\label{se} \delta_{\star} T_K (y) = -\int d^{d}x~\frac{\delta (\delta_{\star} \Phi^{I}(x))}{\delta \Phi^{K}(y)}~T_I(x) . \eea 
This equation represents the condition of \textit{strong invariance} of the equations of motion (EOMs), which is valid off-shell.

If one goes on-shell, ie. imposes $T=0$ and apply to \eqref{se}, then we get only 
\bea{}\label{winv} \delta_{\star} T_K\thickapprox 0. \eea
This equation denotes the \textit{weak invariance} of the EOMs and the symbol `$\thickapprox$' tells us that the statement above is valid only on-shell. 
Weak invariance of EOMs denotes necessary condition whereas strong invariance is considered as a sufficient condition for any generator $\star$ of a given algebra to be a symmetry of the action. 

Let us check whether these equations of motion have strong invariance using the representation of the Carrollian algebra. The equations in magnetic limit are given as
\bes{}\label{cem}
\bea{}\label{cem1}&&T_j:=\p_j\p_t B_t +\p_{t}\p_t B_j=0,\\\label{cem2}
&&T_B:=\p_t\p_j A_j+c_5 \p_t \p_t B_t=0,\\\label{cem3}
&&T_{B_j}:=\p_t\p_t A_j=0.
 \eea\ees
The transformations will be appropriate conformal Carroll transformations corresponding to 
the values: \bea{}\label{reps} (B_t, B_i): \D=1,q=0,q'=-1 ~~\text{and}~~A_i: \D=1,q=-1,q'=0.\eea 

The general expression for strong invariance \eqref{se} for this theory becomes
\bea{}\label{ss}&& \delta_{\star}T_{K} (t,x) = -\int d^{3}y\,dt'~\Big[\frac{\delta (\delta_{\star}A_{i}(t',y))}{\delta \Phi_{K}(t,x)}~T_i(t',y)+\frac{\delta (\delta_{\star}B_t(t',y))}{\delta \Phi_{K}(t,x)}~T_B(t',y)\non\\&&\hspace{6.3cm}+\frac{\delta (\delta_{\star}B_i(t',y))}{\delta \Phi_{K}(t,x)}~T_{B_i}(t',y)\Big].
 \eea
where $\star$ denotes Carrollian conformal generators, $T_{K}=T_j, T_{B_j}, T_B$ and $\Phi_{K}=A_i,B_i,B_t$ respectively. Under dilatation $D$, the left hand side of \eqref{ss} for \eqref{cem} becomes
\bea{}
\delta_{D}T_{K}=[t\p_t +x_l\p_l +3]T_{K}.
\eea
The right hand side of \eqref{ss} for \eqref{cem1} gives,
\bea{}&&
\delta_{D}T_j=-\int d^{3}y\,dt'~\Big[\frac{\delta (\delta_{D}A_{i}(t',y))}{\delta A_j(t,x)}~T_i(t',y)\Big]\non\\&&\hspace{1cm}=-\int d^{3}y\,dt'~\Big[\frac{\delta (t\p_tA_{i}+y_l\p_lA_i+A_i)}{\delta A_j(t,x)}~T_i(t',y)\Big]\non\\&&\hspace{1cm}=[t\p_t +x_l\p_l +3]T_j.
\eea
We see that the \eqref{cem1} have strong invariance under dilatation. Similarly, other equations of \eqref{cem} are strongly invariant under $D$. 

We will now see the strong invariance of \eqref{cem} under $K_l$. The left hand side of \eqref{ss} for \eqref{cem} becomes
\bes{}
\bea{}&&\label{as1}
\delta_{K_l}T_j=(2x_l t\p_t +2x_l x_m \p_m -x^2 \p_l +6x_l)T_j+2x_m\delta_{lj}T_m-2x_jT_l,\\&&\label{as2}
\delta_{K_l}T_B=(2x_l t\p_t +2x_l x_m \p_m -x^2 \p_l +6x_l)T_B+2tT_{B_l},\\&&\label{as3}
\delta_{K_l}T_{B_j}=(2x_l t\p_t +2x_l x_m \p_m -x^2 \p_l +6x_l)T_{B_j}+2x_m\delta_{lj}T_{B_m} \non \\ 
&& \hspace{5cm}-2x_jT_{B_l}.
\eea\ees
The right hand side of \eqref{ss} for \eqref{cem} becomes
\bea{}&& \delta_{K_l}T_j=-\int d^{3}y\,dt'~\Big[\frac{\delta (\delta_{K_l}A_{i}(t',y))}{\delta A_j(t,x)}~T_i(t',y)\Big]=\eqref{as1},\non\\&&
\delta_{K_l}T_B=-\int d^{3}y\,dt'~\Big[\frac{\delta (\delta_{K_l}B_{i}(t',y))}{\delta B_t(t,x)}~T_{B_i}(t',y)\non \\
&& \hspace{2cm}+\frac{\delta (\delta_{K_l}B_{t}(t',y))}{\delta B_t(t,x)}~T_B(t',y)\Big]=\eqref{as2},\non\\&&
\delta_{K_l}T_{B_j}=-\int d^{3}y\,dt'~\Big[\frac{\delta (\delta_{K_l}B_{i}(t',y))}{\delta B_j(t,x)}~T_{B_i}(t',y)\Big]=\eqref{as3}.
\eea
We conclude from the above analysis that the equations \eqref{cem} are strongly invariant under Carrollian generators in $d=4$ dimensions.

We will now look at the strong invariance under infinite Carroll `super-translations' $M_f$. The left hand side of \eqref{ss} for \eqref{cem} becomes
\bes{}
\begin{eqnarray}&&\label{ds1}
\delta_{M_f}T_j =f(x)\p_t T_j,\\&&\label{ds2}
\delta_{M_f}T_B= f(x)\p_tT_B +[\p_jf(x)]T_{B_j},\\&&\label{ds3}
\delta_{M_f}T_{B_j} = f(x)\p_tT_{B_j} .
\end{eqnarray}\ees
The right hand side of \eqref{ss} for \eqref{cem} gives
\bea{}&&\hspace{-.75cm} \delta_{M_f}T_j=-\int d^{3}y\,dt'~\Big[\frac{\delta (\delta_{M_f}A_{i}(t',y))}{\delta A_j(t,x)}~T_i(t',y)\Big]=\eqref{ds1},\non\\&&\hspace{-.75cm} 
\delta_{M_f}T_B=-\int d^{3}y\,dt'~\Big[\frac{\delta (\delta_{M_f}B_{i}(t',y))}{\delta B_t(t,x)}~T_{B_i}(t',y)+\frac{\delta (\delta_{M_f}B_{t}(t',y))}{\delta B_t(t,x)}~T_B(t',y)\Big]=\eqref{ds2},\non\\&&\hspace{-.75cm} 
\delta_{M_f}T_{B_j}=-\int d^{3}y\,dt'~\Big[\frac{\delta (\delta_{M_f}B_{i}(t',y))}{\delta B_j(t,x)}~T_{B_i}(t',y)\Big]=\eqref{ds3}.\non
\eea
\textit{We therefore confirm the strong invariance for \eqref{cem} under the infinite dimensional conformal Carrollian symmetry algebra in $d=4$ and these are true dynamical symmetries.}

Finally, we can write down the Lagrangian which gives the equations of motion \eqref{rer}. It is given by
\begin{eqnarray}\label{l0}
L_0= \int d^3x\Big[(\partial_j A_j)( \partial_t B_t) + (\partial_t A_j)( \partial_t B_j )+ \frac{c_5}{2} (\partial_t B_t)^2 \Big].
\end{eqnarray}
We note here that for all the dynamical variables in the above Lagrangian, all time derivatives can be uniquely be solved in terms of the canonical momenta and the Hessian is invertible. Hence, according to the Dirac prescription, \textit{the system is free of constraints and hence devoid of gauge redundancy}! Ones perceive it in a way that the ultra-relativistic limit breaks the $U(1)$ gauge invariance at the level of Lagrangian. This is in contrast to viewing Maxwell theory as a massless limit of Proca theory. In Proca dynamics, gauge invariance (as well as scale invariance) emerges as $m \rightarrow 0$. In contrast here, we have a situation where gauge invariance gets broken as speed of light $c \rightarrow 0$. However the interplay between ultra-relativistic limit and the issue of gauge is more subtle, which we will discuss later in the next section.

We notice here a particular cross kinetic term $\partial_t A_j ~ \partial_t B_j$. From the perspective of a quantum theory this does not cause any problem, if one is interested to extract correlation functions from a formally defined path integral \footnote{Notably, for the world-sheet way of looking at string theory with Lorentzian target space-time, there is always terms with `wrong' kinetic term. However that does not prevent one from constructing a unitary theory with physically meaningful spectrum.}. This can be achieved by first complexifying the space of fields and then defining the path integral contour such that the Gaussian determinant is well defined \cite{Witten:2010zr}.  Although we don't attempt to compute the determinant for the free theory as that would just give an adjustable normalization factor for loop computations in an interacting theory, in order to facilitate that for a future progress, we exemplify one such analytic continuation (among various possible others) below.

Let us first promote the real vector fields $A_i, B_i$ to complex ones. Then we reparametrize the fields as:
\begin{equation}\label{eded}
 A_i = \frac{1}{2} (D_i + iE_i), \quad B_i = \frac{1}{2}(D_i -i E_i). \;
\end{equation}
We now recast the Lagrangian \eqref{l0} of our theory in terms of fields $ D_i $ and $ E_i $ and add a complex conjugate term to make it real:
\bea{}\label{lklk}
\tilde{L}&=&\int d^{3}x\,\Big[\frac{1}{2}( \partial_t B_t)\{\partial_j D_j+ i\partial_j E_j \}+\frac{1}{4}( \partial_t D_j )^2 +\frac{1}{4}(\partial_t E_j)^2 + \frac{c_5}{2} (\partial_t B_t)^2 \non \\  && +
\mbox{ complex conjugate } \Big].
\eea
With the above prescription of analytic continuation, calculation of the partition function and hence the correlators calculated from it make perfect sense. Secondly as mentioned above, the system is completely devoid of gauge invariance and hence correlators calculation from the partition functions is unambiguous. From the perspective of global symmetries, we have a unique vacuum as defined in section 2.2. Therefore, the correlators  discussed there would be the same as obtained from \eqref{lklk}. Hence we can attempt to make a curious connection with the correlation functions in the original Maxwell theory, which are of course heavily dependent upon choice of gauge. 
Following the results in the table [\ref{tab:2pt}], we have:
\begin{table}[H]
\begin{center}
\begin{tabular}{ |p{.5cm}|p{4cm}|p{4cm}|}
 \hline 
& Correlators &Results\\
 \hline
 1.  &$\langle B_t (t_1, x_1) B_t (t_2, x_2) \rangle$&$=0$\\
2.& $ \langle B_t (t_1, x_1)D_j (t_2, x_2)\rangle$  & $=0$\\
3. &$  \langle B_t (t_1, x_1)E _j (t_2, x_2) \rangle$ & $=0$\\
\hline\hline
4.  &$\langle D_i (t_1, x_1) D_j (t_2, x_2) \rangle$&$=\frac{a}{r^2} \delta_{i j}+ \frac{b}{r^4} x_i x_j$\\
5.&  $ \langle D_i (t_1, x_1)E_j (t_2, x_2)\rangle$  & $= \frac{c}{r^2} \delta_{i j} +\frac{d}{r^4}x_i x_j$\\
6. &$  \langle E_i (t_1, x_1)E _j (t_2, x_2) \rangle$ & $=-\frac{a}{r^2} \delta_{i j}- \frac{b}{r^4} x_i x_j$\\
\hline 
\end{tabular} 
\caption{Summary of results}\label{tab:jk1}
\end{center}
\end{table}
\begin{figure}[h]
\centering
\includegraphics[scale=0.3]{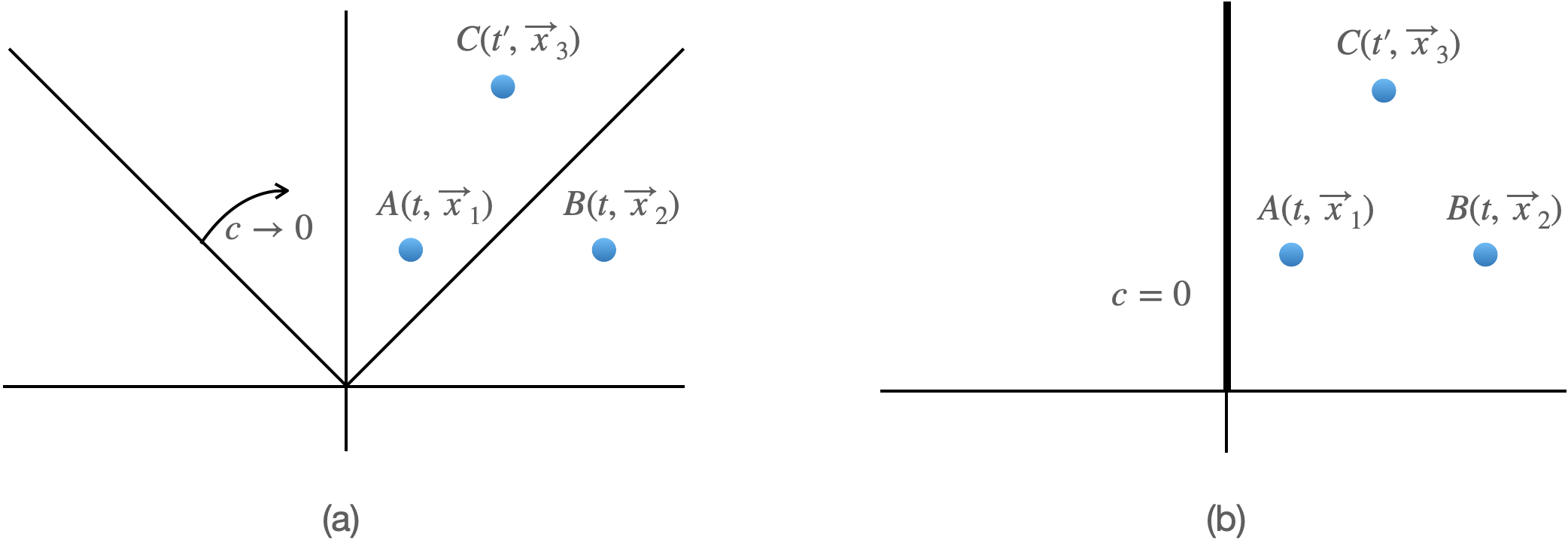} 
\caption{In the panel (a) above, the points $A$ and $C$ are causally connected, whereas the interval $A$ - $B$ is acausal, in Minkowski space. To get Carroll space-time as the $c \rightarrow 0$ limit, one should collapse the light cone to a single line, as in panel (b). In Carroll space-time, none of the points $A, B$ and $C$ are causally connected, unless the events take place exactly at the same spatial points.}
\label{fig:causal}
\end{figure}
In Appendix [\ref{asasasa}], we kept for a comparison, the causal interval $(|x^0-y^0|>|\vec{x}-\vec{y}|)$ as well the acausal $( |\vec{x}-\vec{y}|>|x^0-y^0|)$ ones for photon propagator in a 1-parameter family of gauge choices. Being a massless free theory of long-range interactions, for acausal intervals the propagator has $1/r^2$ spatial fall-off. Before thinking about a Carrollian limit of the photon propagators, we notice that same manifold points which are causally connected in Minkowski space-time, are no more so in Carrollian limit unless they are at coincident spatial points (See Figure [\ref{fig:causal}]). 
Now, two acausal events can be made to occur in common time, by a continuous Lorentz transformation and similarly two causally connected events can be brought to occur at a coincident spatial point. In this sense, it is natural that we take a Carrollian limit of acausal photon propagator in a particular gauge to compare with our Carrollian propagators in Table [\ref{tab:jk1}]. Curiously, the results mentioned in Table [\ref{tab:title3}], at the ultra-relativistic limit can be compared with the ones given in Table [\ref{tab:jk1}]. If we take the gauge fixing parameter $\xi=-3$ there, the correlator between temporal components vanish, as in the table above. On the other hand the structure of the correlators between spatial components also match exactly. 

Motivated by this, we check in the next subsection whether the free Carrollian theory we have, indeed comes from an ultra-relativistic limit of any Lorentz covariant theory of two vectors.

\section{Looking for a Minkowski ascendant}
The Lagrangian $L_0$ obtained above in \eqref{l0} have equations of motion which correspond to the magnetic limit of
Carrollian electrodynamics once the additional fields $B_i$ and $B_t$ are set to zero. However, it is not clear whether
the entire theory by itself can be obtained by taking suitable limits of a Lorentz invariant theory\footnote{RB thanks Andrew Strominger for suggesting this check}. In this section, we will try to answer this question. 

\begin{figure}[h]
\centering
\includegraphics[scale=0.3]{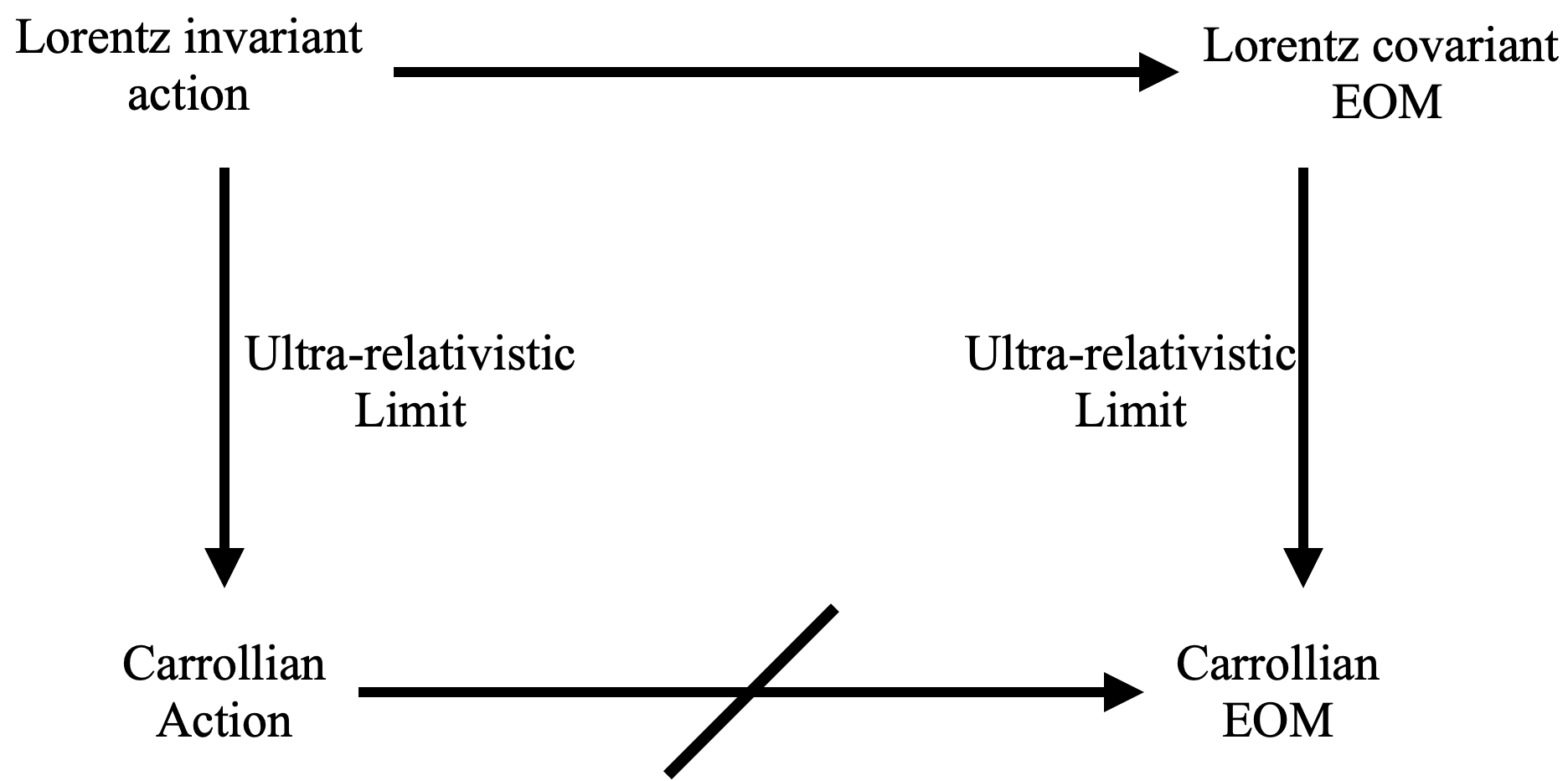} 
\caption{The process of taking Carrollian (ultra-relativistic) limit and applying variational derivative on action functional do not commute.}
	\label{fig:noncommute}
\end{figure}

Recall that the equations of motion for the theory given in \eqref{cem} contains two $SO(3)$ vectors and a scalar. Hence, if the theory is to descend from a Minkowski theory as an ultra-relativistic limit, the Lorentz covariant theory better have a couple of Lorentz vectors, giving rise to a couple of $SO(3)$ vectors and scalars. For degree of freedom matching one scalar must be rendered unphysical or non-dynamical. One way to see if such a parent theory exists, is by starting with a Lorentz invariant action with two Lorentz covariant vectors and then taking limits. However, we already noted in section 3.1, 
the diagram in Figure [\ref{fig:noncommute}] does not necessarily commute.

Hence we start with the most general \eqref{most_general} Lorentz covariant equations of motion and take the ultra-relativistic limit and impose Helmholtz condition (such that there exists an Carrollian action). Then we inspect whether making one $SO(3)$ scalar non-dynamical would yield us \eqref{cem}.

We begin by writing out a set of most general Lorentz covariant equations of motion for two vector fields (say $ A_{\mu
}$ and $ B_{\mu}$). ie,
 \begin{eqnarray}\label{most_general}
   T_A &=& a_1 \partial_\nu \partial^\nu A_\mu + a_2 \partial_\mu \partial_\nu A^\nu + a_3 \partial_\nu \partial^\nu B_\mu + a_4 \partial_\mu \partial_\nu B^\nu = 0, \nonumber \\
   T_B &=& b_1 \partial_\nu \partial^\nu A_\mu + b_2 \partial_\mu \partial_\nu A^\nu + b_3 \partial_\nu \partial^\nu B_\mu + b_4 \partial_\mu \partial_\nu B^\nu = 0.
  \end{eqnarray}
There's a caveat regarding gauge, in the equations presented above and a bit of care is necessary before one takes Carrollian limits on them. That is, unless we choose particular constraints on the coefficients $a_1, a_2$ etc., the equations don't enjoy gauge invariance. However if one puts a constraint $a_1 = - a_2$, then  terms like $\p^2_t A_t$ drop out from \eqref{most_general}. Therefore taking Carrollian limits on those equations would lead us to equations of motion which describe completely different physics than the case of $a_1 \ne - a_2$.

Hence before ultra-relativistic limits, we classify all such classes of possible gauge (keeping in mind that the equations of \eqref{most_general} should not be degenerate in any case) invariance.

\begin{enumerate}
\item All the coefficients $a_i, b_i, \, i =1, \dots 4$ are non-zero and independent of each other.
\item $a_1 = -a_2$ and all the other parameters independent of each other. In this case the first of the equations \eqref{most_general} has gauge invariance in the field $A_\mu$. However the second one breaks it.
\item $a_1 = -a_2$ and $b_1 = -b_2$ and all other parameters are independent. Therefore both the equations is $U(1)$ gauge invariant for the field $A$.
\item $a_1 = -a_2$ and $a_3 = -a_4$ and all other parameters are independent. The first equation has $U(1) \times U(1)$ gauge invariance whereas the second equation breaks both.
\item $a_1 = -a_2$ and $b_3 = -b_4$ and all other parameters are independent. Both the equations separately is $U(1)$ gauge invariant but those are broken by respectively the other one.
\item $a_1 = -a_2, a_3=-a_4$ and $b_1 = -b_2$ and other parameters are independent. In this case the first equation is $U(1) \times U(1)$ gauge invariant but the second one looses half of it.
\item $a_1 = -a_2, a_3 =-a_4, b_1 = -b_2$ and $b_3 = -b_4$. The system describes fully $U(1) \times U(1)$ gauge invariant fields.
\end{enumerate}
In each of these above classes, one can take at most 4 combinations of Carrollian electric and magnetic limits on the $A_{\mu}$ and $B_{\mu}$ fields and finally set $A_t$ or $B_t$, the $SO(3)$ scalars to space-time constant. It's a straightforward yet strenuous set of calculations and then comparing with \eqref{cem} which we performed. \textit{But none of the above equations result into the Carrollian magnetic equations of motion \eqref{cem}.} 
Hence we conclude that there is no Lorentz invariant theory whose ultra-relativistic limit is the proposed magnetic sector of Carrollian electrodynamics.

However, the possibility remains that our theory can be obtained from a gauge fixed version of a Minkowski theory, where the gauge fixing condition explicitly breaks the Lorentz symmetry. This can be achieved by setting some of the $a_i$ or $b_i$ arbitrarily to zero or some other fixed number.
Such a possibility cannot be explored in this analysis carried out in this section but we briefly explore a way to address this issue in 
Appendix [\ref{asasasa}].

\section{Extension to an interacting theory}
Until now, we have been concerned with Carrollian version of electrodynamics, which is a free theory enjoying an infinite dimensional global symmetry group. Although not very common, it's understandable that a free theory might have much extended symmetry, hidden or explicit (view Appendix \ref{infu} for such a class of symmetries). However, in the following we will present very non-trivial result in the form of interaction terms, such the new interacting theory still enjoys the infinite dimensional conformal Carrollian symmetry.

\subsection{Addition of Interactions}\label{aoi}
One way of adding interactions to electrodynamics is to couple minimally charged matter. Another would be to generalize to non-abelian version of it. Both of these procedures rely heavily on the gauge principle, which the Carrollian theory lacks.

Our principle in finding possible deformations to the free theory starts by first classifying all possible marginal terms (hence respecting scaling symmetry) which respect spatial rotational symmetry. This would involve all possible $SO(3)$ scalars of dimension 4. We first exhaust all possible terms of this type. This includes, for example, $B^4_t, D_i D_i E_j E_j$ or momentum dependent vertices like $B^2_t \partial_j D_j$ etc. We add all such interactions and then check for invariance under the infinite dimensional conformal Carrollian symmetry. Through a lengthy yet straightforward set of computations, directly checking with the symmetry generators presented in Table [\ref{tab:cvc}], we see that the Lagrangian is invariant upto total time derivative, only if we include the following terms to the free Lagrangian \eqref{lklk}
\begin{eqnarray}
\tilde{L}_{int} = \int d^3x\Big[-g_1 B_t^4 - 
\frac{g_2}{2} B_t^2 (D_j^2-E_j^2)\Big].
\end{eqnarray} 
Therefore, symmetry principle (in particular the infinite dimensional conformal Carrollian algebra) helps us find the interactions uniquely. Hence we have constructed an example of an interacting theory which possess the infinite number of conformal Carrollian global symmetries.

\subsection{Dynamical Realization of the Carrollian algebra}
The conformal Carroll algebra is not semi-simple and has an infinite dimensional abelian ideal. For non semi-simple symmetry algebras, due to cohomological properties, it is not guaranteed that the homomorphism from the algebra of symmetry generators to the space of conserved charges is exact \cite{Woodhouse:1980pa}. This gives rise to the possibility of non-trivial central extension. The easiest example is that of a 1-D free particle Lagrangian. This has Galilean boost and translation as symmetries. The two dimensional symmetry algebra is abelian and not semi-simple, obviously. It's second cohomology is non-trivial. This is reflected in the mass dependent central term in the Poisson bracket between conserved momentum and boost charge\footnote{The author RB thanks Glenn Barnich for brining to notice this naive yet beautiful example. A more involved but standard example where Noether charge algebra does not exactly respect the algebra of symmetry generators (central term appears), crops up in the asymptotic symmetries of asymptotically flat space-time.}. It is therefore instructive to check for possible central extensions in our present case of interest.

For the interacting theory \eqref{lklk}, we will now look at the Noether charges and the corresponding Poisson brackets.
Although our Lagrangian is manifestly invariant under CCA, the closure of the algebra of Noether charges
will ensure the dynamical preservation of the Conformal Carrollian symmetries. Let us consider the conserved Noether charges consistent 
with the symmetries associated with the Lagrangian.
Since we don't have manifested Lorentz covariance in our system, calculating directly the textbook definition of Noether
current and thereafter the Noether charge would not be feasible. 

The systematic procedure we employ to find out the charges is as follows: Consider a Lagrangian in $d$ spacetime dimensions 
\bea{}
L=L(\Phi,\p_t \Phi, \p_i \Phi ).
\eea
where $\Phi(t,x)$ is a generic field. Varying the Lagrangian on-shell in an arbitrary direction on the tangent space of field space: $\Phi \rightarrow \Phi +\delta \Phi$ , we get
\bea{}\label{theta}
\delta L= \int d^{d-1}x \:\Big[\p_t \underbrace{\Theta(\Phi, \p \Phi, \delta \Phi)}_{\mbox{(pre)-symplectic potential}} \Big] ~~: \text{on-shell.}
\eea
Now consider a specific infinitesimal transformation  $\Phi \to \Phi +\delta_{\star} \Phi$ off-shell. The transformation $\delta_{\star}$ is said to be a symmetry, if:
\bea{}\label{alpha}
\delta L= \int d^{d-1}x \:\Big[\p_t f  (\Phi, \p \Phi, \delta_{\star} \Phi) \Big] ~~: \text{off-shell,}
\eea
for some function $f$ in field space.

Comparing \eqref{theta} and \eqref{alpha}, we infer that on-shell:
\bea{}\label{inter}
\p_t Q_{\star} := \int d^{d-1}x \p_t \left( \Theta(\Phi, \p \Phi, \delta_{\star} \Phi) - f  (\Phi, \p \Phi, \delta_{\star} \Phi) \right) =0
\eea
Using this procedure, the Noether charges for the finite and infinite Conformal Carrollian generators are calculated as
\bea{}\label{charcan1}
&&\non \hspace{-.5cm}\text{Rotation:}~~  Q_{\omega}=\int d^{3}x\  2 \omega^{i j} \Big[  x_i \{2 \pi_{D_l}\p_jD_l+ 2 \pi_{E_l}\p_jE_l
+2 \pi_B \partial_j B_t \} \non\\&&\non\hspace{7.5cm}+2 \pi_{D_i}D_j +2 \pi_{E_i} E_j\Big],\\&&\non\\&&\non\hspace{-.5cm}
\text{Translation:}~~  Q_{p}=\int d^{3} x \ p^{k}\Big[ 2 \pi_{D_j}\p_kD_j + 2 \pi_{E_j}\p_k E_j+
+ 2\pi_B \partial_k B_t \Big],\\&&\non\\&&\hspace{-.5cm}
\text{ST:}~~Q_{f}=\int d^{3} x \, f\Big[ 2 \pi_{D_j}^2+2 \pi_{E_j}^2+\frac{1}{c_5}\Big(\pi_B -\frac{1}{2}( \partial_j D_j+i\partial_j E_j) \Big)^2\non \\&&\hspace{3cm}
+\frac{g_2}{2}B_{t}^{2} (D_j+i E_j)^2
+2g_1B_{t}^{4}\Big\}+ 2 \p_j (B_t(\pi_{D_j}+i\pi_{E_j}))\Big],\non\\&&\non\\&&\hspace{-.5cm}
\text{Dilatation:}~~ \non Q_{D}=\int d^{3} x  \,\Big[2\pi_{B} B_{t}+(2 \pi_{D_{j}}D_j+ 2 \pi_{E_{j}}E_j)
+2 x^{l}\{\pi_{B}(\partial_l B_{t})\non\\&&\hspace{4.5cm}+\pi_{D_{j}}\p_lD_j+\pi_{E_{j}}\p_lE_j\} 
+ 2 t\Big\{\pi_{D_j}^2+\pi_{E_j}^2 \non\\&&\hspace{4.5cm}+\frac{1}{2 c_{5}}\Big(\pi_{B}-\frac{1}{2}(\partial_{l} D_{j}+i \partial_{l} E_j)\Big)^2 
+g_{1} B_{t}^{4}\non\\&&\hspace{4.5cm}\non\\&&\hspace{4.5cm}+\frac{g_{2}}{4} B_{t}^{2}\Big(( D_{j}+iE_{j})^{2}\Big)\Big\}\Big].
 \eea
 Similarly, the charge associated to special conformal transformation is given as 
\bea{}\label{cds}&&\non\hspace{-.6cm}
Q_k= \int d^{3} x \  2k_{l} \Bigg\{(D_l +iE_l)B_{t} +2 x_{l} \Big[\pi_B B_t+\pi_{D_j}D_j+\pi_{E_j}E_j  
 \Big]\non\\&&\hspace{2.1cm}+ 2 x_{l} t \Big[\pi_{D_j}^{2}+\pi_{E_j}^{2}+
+\frac{1}{2c_5}\Big(\frac{1}{2}( \partial_{j} D_{j}+i \partial_{j} E_{j})-\pi_B \Big)^2 +2g_{1} B_{t}^{4} \non\\&&\hspace{2.1cm}+\frac{g_2}{4} B_{t}^2\Big(D_{j}+i E_{j}\Big)^2 \Big] \non\\&&\hspace{2.1cm}
+x_{l} x_{m}\Big[ 2\pi_{B}(\partial_m B_{t})+2 \pi_{D_j}\p_mD_j+2 \pi_ {E_j}\p_mE_j\Big] \non\\&&\hspace{2.1cm}
-x^2\Big[\pi_{B}(\partial_l B_{t})+ \pi_{D_j}\p_lD_j+ \pi_ {E_j}\p_lE_j\Big] \non\\&&\hspace{2.1cm}
+ 2 x_{m}\Big[ (\pi_{D_l}D_m+ \pi_ {E_l}E_m\Big] \non\\&&\hspace{2.1cm}
- 2 x_{j}\Big[ ( \pi_{D_j}D_l+ \pi_ {E_j}E_l \Big] \non\\&&\hspace{2.1cm}
- 2t( \pi_{D_l}+i \pi_{E_l})B_t \Bigg\}.
\eea
We will write down the results of Poisson brackets between the conserved charges using the canonical commutation
relations. The Poisson bracket between dilatation and spatial translation is given by
\be{al1}
\{Q_D,Q_p\}=-Q_p.
\ee

The above Poisson brackets \eqref{al1} reflects the CCA bracket
\be{}
\lb D,P_k\rb=-P_k.
\ee
Consider some of the other terms in the infinite CCA:
\bea{}
&&\non [P_k, M_f] =M_{\p_k f},\quad  [D,M_f] =M_h,~\text{where}~h=x_l \p_l f-f,\\
&&\non [K_l,M_f]= M_{\tilde{h}},~\text{where}~\tilde{h}=2x_l h-x^2\p_l f.
\eea
Correspondingly, a set of lengthy yet straightforward Poisson bracket computations of the charges give:
\bes{}\label{caa}
\bea{}
&&\{Q_p,Q_f\}=Q_{h^\prime},~~\text{where,~}h^\prime=p^k \p_k f,\\
&&\{Q_D,Q_f\}=Q_h,~~\text{where,~} h=x_k\p_k f-f,\\
&&\{Q_k,Q_f\}=Q_{\tilde h},~~\text{where,~}\tilde h=(2x_i x_k \p_k-x^2\p_i)f.
\eea
\ees
The results confirm the CCA algebra being satisfied at the level of charges, ie the existence of Lie algebra homomorphism from CCA to the Poisson algebra of charges. Given the fact that CCA is infinite dimensional and non-semi simple, this check was necessary. Because for non-semi simple symmetry algebras, this homomorphism is not guaranteed and may lead to central extensions.
\section{Conclusions}
\subsection*{Summary}
To conclude, let us first summarize the results obtained in the paper.

Our main achievement in this paper has been the construction of an interacting theory with infinite number of global symmetries in $d=4$. Starting from the magnetic sector of ultra-relativistic equations of motion, we added newer degrees of freedom to the system to make it dynamically consistent. The resultant theory is devoid of gauge redundancies. And more interestingly, addition of new degrees of freedom takes us to a portion of the space of Carrollian theories which are not, in an obvious way, found to be ultra relativistic limit of any Lorentz invariant field theory.

To begin with, we started with a brief review of flat conformal Carrollian manifolds and isometries thereon, which form an infinite dimensional lie group. For fields, which we demand transform as scalars and vectors under spatial rotations and to have definite scaling dimensions, we develop the transformation rules under the conformal Carrollian transformations. We then motivate towards a theory of Carrollian electrodynamics as an ultra-relativistic limit of Maxwell's electrodynamics and focus particularly on the magnetic sector, at the level of equations of motion. As these equations of motion can't be derived as Euler-Lagrange equations from an action, we device an algorithm for adding newer degrees of freedom linearly to the system respecting the symmetries, so that we have an action principle to work with. Once we have an action, we add interaction terms which such the system has infinite dimensional conformal Carroll group as group of global symmetry generators.

\subsection*{Discussions and Future Direction}
Following are some of the aspects of present article which should be pursued in the near future.

\medskip

\noindent \textit{Propagators}

\smallskip

\noindent As mentioned in the introduction, when viewing Carrollian space-time as an ultra-relativistic limit of Lorentz covariant systems, light-cones now collapse to the erstwhile time axis of Minkowski space. This obviously means that all causal propagation are ultra-local in space. Without going into the picture of particles as quanta of the fields in Carroll background, we worked here with a very mild assumption of existence of a vacuum state of the free theory. With a set of symmetry considerations, including the conformal ones, we are able to construct uniquely the position space propagator. The propagators are unique, even for massless vector theories, as this is devoid of gauge invariance. Interestingly these two point functions can be interpreted as ultra-relativistic limit of relativistic $U(1)$ gauge theory at a certain gauge. Further investigations in causal propagators are necessary to set up the quantum theory. A rather intriguing question in this regard, for those Carrollian theories which descend as ultra-relativistic limit is whether the connection carries over at the quantum levels as well.

\medskip

\noindent \textit{Perturbative Quantization}

\smallskip

\noindent One of the key motivations in this line of projects in Carrollian physics is probing and classifying all CFTs,
beyond the regime of the relativistic ones and other QFTs connected via RG flow. Taking cue from relativistic physics,
we have included all possible marginal deformations in the present theory. So it is imperative that one should check the
divergence structure at least up to the first quantum correction and understand the meaning of renormalization in Carrollian set up.

\medskip

\noindent \textit{Ward identities}

\smallskip

\noindent Probably the main feature that sets Carrollian conformal theories apart is the existence of an infinite set of locally acting space-time symmetry transformations. 
If a consistent quantization program is developed, the obvious question that one would like to ask is the exactness or corrections to these infinite symmetries at quantum level 
via Ward identities. Even finding anomaly structure for the scaling symmetry itself would be an interesting progress in these theories.

\medskip

\noindent \textit{Graphene superconductivity}

\smallskip

\noindent It is well known that low lying levels of electron energy bands in Graphene exhibits linear dispersion, and hence the low energy (comparing to Fermi level) physics is described by Dirac equation in 2+1 dimensions. Hence this is a massless Lorentz covariant description, with speed of light being replaced by the Fermi velocity. It's been recently observed \footnote{We thank Gregory Tarnopolsky for bringing this to notice and Swastibrata Bhattacharyya for further discussions on this topic.} experimentally \cite{Cao:2018} that at certain twist angles (magic angles), the effective Fermi velocity goes to zero and the conical bands flatten out making way of a new type of superconductivity. This phenomena obviously is an indicator of Carrollian conformal physics in Fermionic systems. We would like to explore more into these systems with particular emphasis on possibility and consequences of  Carrollian conformal symmetries.
\section*{Acknowledgement}
It’s a pleasure to thank Arjun Bagchi for collaboration at the initial stage of the work and further numerous discussions and inputs, who is also instrumental in developing an ongoing program on Carrollian field theories.

\noindent We acknowledge Andrew Strominger and Max Riegler for useful comments and discussions. RB gratefully
acknowledges Fulbright Foundation, DST, India Inspire Faculty grant and SERB for support through SRG/2020/001037 grant. AM acknowledges Alexander von Humboldt
Foundation for the Humboldt Research Fellowship. KB and RB thanks SERB CRG/2020/002035 for support. 
\appendix
\section{Digression on Infinite Dimensional Symmetry Algebras in Relativistic theories} \label{infu}

It is not often realized that even (free) field theories in Minkowski space have infinite number of continuous symmetries and hence conserved charges. In the following, we will explore a large set of them.

This is motivated basically by an algebra of abelian generators, similar in spirit to the supertranslation charges $Q_f$ of BMS symmetries, ie. those forming the ASG of general relativity for asymptotically flat spacetime. As a theory of gravity must have built in diffeomorphism invariance, any non-trivial physical symmetry and hence conserved quantity is only supported at the asymptotic boundary. For asymptotically flat spacetimes, $Q_f$ charges are defined as conserved charges integrated over 2-sphere foliations of the future (or past) null infinity, corresponding to arbitrary angle dependent time translations, off the celestial sphere. Since $f$ is an arbitrary function, there are infinite number of them, which are algebraically independent and by construction, they are all conserved global charges. Note that, the energy $H = Q_{f=1}$ is a special case of the supertranslation charge. 

In order to stress on the existence of the non-triviality of the infinite number of algebraically independent conserved quantities in 4 dimensional bulk physics (free), we bring in the textbook topic of a field theory defined on a Lorentzian manifold like Minkowski, where one can still define phase space functions like 
\bea{}
Q^{\mathrm{(Mink)}}_f = \int d^3 x \,f\, \mathcal{H}^{\mathrm{(Mink)}}
\eea
with $\mathcal{H}^{\mathrm{(Mink)}}$ being the Hamiltonian density for the relativistic theory on Minkowski space and the function $f$ is supported only on the spatial surface. For local, Poincare invariant theories\footnote{eg. for a massless free scalar in $3$ spatial dimensions, 
$\mathcal{H}^{\mathrm{(Mink)}} =\frac{1}{2} \left( \pi^2 + \p_i \phi \p_i \phi\right)  $ and hence,
\bea{commi}
\frac{d}{dt} Q^{\mathrm{(Mink)}}_f  = -\int d^3 x \, \p_i f \, \p_i \phi \, \pi. \non
\eea 
 } of course these won't be conserved unless $f$ is constant. 
 
However, since going to the momentum space completely decouples free theories as independent oscillators, an infinite number of conserved quantity can be constructed. 
In order to facilitate the comparison, one can start with the Hamiltonian of a massless free field theory of helicity $\sigma$ (taken to be integral; otherwise one has to take a  little bit more care in the following discussion the variables now becoming Grassmann):  
\bea{shamik_ham}
H^{\mathrm{(Mink)}} = \frac{1}{2(2\pi)^3}\int d^3 \vec{p} \,|\vec{p} | \, a^{\star} (p, \sigma) a (p, \sigma)
\eea
with the usual (pre)-symplectic structure:
\bea{shamik_symp}
\Omega = -\frac{i}{2(2 \pi)^3} \int d^3 \vec{p} \, \mathbb{D}a (p, \sigma) \wedge \mathbb{D} a^{\star} (p, \sigma). 
\eea
Here 
\bea{nullm}
p = \{ (E, \vec{p}) \Big| E^2 - |\vec{p}|^2 = 0\}\eea
 is the null momentum and we denote the exterior derivative on phase space by $\mathbb{D}$. Note that, we have not chosen a traditional Lorentz invariant measure in \eqref{shamik_ham} and density factors have been appropriately absorbed in oscillator variables, which is reflected in the symplectic structure. 

 It is easy to verify that the vector field,
 $$ \xi =  i \int d^3 {\vec{p}}'  \left( a(p', \sigma) \frac{\delta}{\delta a(p', \sigma)} - {a^{\star}}(p', \sigma) \frac{\delta}{\delta {a^{\star}}(p', \sigma)} \right)$$ on the infinite dimensional phase space is a generator of canonical transformation and the corresponding generating function is the Hamiltonian \eqref{shamik_ham} itself, ie: $i_{\xi} \Omega = \mathbb{D} H^{\mathrm{(Mink)}}$. It captures the time translation symmetry of the problem.

Interestingly, a phase space vector field 
\bea{xig}
\xi_g =  i \int d^3 {\vec{p}}'  g({\vec{p}}') \left( a(p', \sigma) \frac{\delta}{\delta a(p', \sigma)} - {a^{\star}}(p', \sigma) \frac{\delta}{\delta {a^{\star}}(p', \sigma)} \right)  
\eea
is also a generator of canonical transformation, for any arbitrary (tensor) $g$  of $\vec{p}$, giving rise to the generating function:
\bea{Qg} Q_g = \frac{1}{2(2\pi)^3}\int d^3 \vec{p}  \, g( \vec{p})\, a^{\star} (p, \sigma) a (p, \sigma). \eea
The $g = |\vec{p}|$ case corresponds to the Hamiltonian \eqref{shamik_ham}. Moreover, these are all conserved:
\bea{conservation} \{ H, Q_{g}\} =  i_{\xi_g} i_{\xi_{g=|\vec{p}|}} \Omega = 0.
\eea
Since $g$ is arbitrary, we already get an infinite number of which are all conserved and they form the infinite dimensional	 Abelian algebra (following the algebra of the symplectomorphisms \eqref{xig}):
$$\{ Q_{f}, Q_{g}\} =  i_{\xi_g} i_{\xi_f} \Omega = 0. $$
In the analysis of finding the BMS symmetry algebra in free field theories in \cite{Banerjee:2018fgd} \footnote{It is to be noted that in \cite{Banerjee:2018fgd}, the functions $g$ were chosen to be supported on the 2-sphere parametrized by $\vec{p}/|\vec{p}|$.}, these generators took the role of supertranslations. Physically this is another manifestation of the fact that the energy of each individual oscillator mode (described by $a, a^{\star}$) are conserved independently, as one can choose a particular momentum $\vec{p}_0$ and the energy of the oscillator corresponding to it is found by choosing $g(\vec{p}) = \delta^3 (\vec{p} - \vec{p}_0)$ in \eqref{Qg}.

What might have been overlooked in recent relevant literature is that, these sets of generators are the special cases of a much larger tensor algebra with varying degrees of locality in momentum space. With $SO(3)$ tensors $F,G$ one can have generators of linear symplectic transformations:
\bea{gen}
&&\int d^3 {\vec{p}} \, F^{i_1 \dots i_m} (\vec p) \frac{\p}{\p p^{i_1}}\dots \frac{\p}{\p p^{i_m}} a(p, \sigma) \frac{\delta}{\delta a(p, \sigma)} \nonumber \\ 
&+&  \, \, G^{j_1 \dots j_n} (\vec p) \frac{\p}{\p p^{j_1}}\dots \frac{\p}{\p p^{j_n}} a^{\star}(p, \sigma) \frac{\delta}{\delta a(p, \sigma)} +  \mbox{c.c.}
\eea
For certain specific symmetry structures and divergence conditions on $F, G$, these also are symmetry generators, called higher spin symmetries for obvious reason. \textit{This is precisely the reason that all free systems are integrable.} Some of interesting exemplary special sub-algebras are as follows.
\begin{itemize}
\item Let's consider the phase-space vector fields:
\bea{}
\chi_{\vec{A}} =   \int d^3 {\vec{p}} \, A^i(\vec{p}) \left( \p_i a(p, \sigma) \frac{\delta}{\delta a(p, \sigma)} + \p_i a^{\star}(p, \sigma) \frac{\delta}{\delta a^{\star}(p, \sigma)} \right).
\eea
This is both a symplectomorphism as well as a symmetry generator, for an arbitrary divergence-less vector field, $\vec{A}$ in 3-momentum space:
\bea{chank}&&\hspace{-.75cm}
 i_{\chi_{\vec A}}\Omega=\mathbb{D}\left(Q[\chi_{\vec A}]\right),  \mbox{ where }\, Q[\chi_{\vec A}] = \frac{i}{2(2\pi)^3}\int d^3\vec{p}\, a^{\star}(p, \sigma) A^i ( \vec{p}) \p_i a (p, \sigma) \nonumber \\&&
 ~\mbox{and}~ i_{\chi_{\vec A}} i_{\xi_{g=1}} \Omega = 0.
\eea

They generate the following Lie algebra of divergence-less vector fields: 
\bea{}
[ \chi_{\vec A}, \chi_{\vec B} ] = \chi_{\vec C} , \, \mbox{ where }\, \vec{C} = \pounds_{\vec B} \vec{A}.
\eea
As expected, this is realized at the level of charges, in an equivariant way:
\bea{}
 i_{\chi_{\vec A}} i_{\chi_{\vec B}} \Omega = Q[\chi_{\vec C}], \, \mbox{ where }\,  \mathbb{D}\left(Q[\chi_{\vec C}]\right)  = i_{\chi_{\vec C}} \Omega.
\eea
\item One can easily parametrize the null 4 momenta forming the null-cone \eqref{nullm} as:
\bea{}
p ^{\mu} = E \left( 1, \frac{z+\bar{z}}{1+z\,\bar{z}}, -i  \frac{z-\bar{z}}{1+z\,\bar{z}}, \frac{1-z\bar{z}}{1+z\,\bar{z}}\right).
\eea
This space with topology ($E >0$) $\mathbb{R}^+ \times S^2$ does not have a Riemann structure, but has conformal properties. Actually, as we will later review in this article, this has properties of a Carroll manifold. One has the induced measure:
$d^3 \vec{p} = i \frac{dE \, dz\, d\bar{z}}{(1+z \bar{z})^2}$. It can be shown that:
\bea{}
L_m = \int_{\mathbb{R}^+ \times S^2}  \frac{dE \, dz \, d\bar{z}}{(1+z \bar{z})^2}\, z^{m+1}  a^{\star}(E,z, \bar{z})\p_z a(E,z, \bar{z})
\eea 
for integer $m$ and analogously written $\bar{L}_n$ are also conserved quantities and their Poisson algebra defined by the above symplectic structure form a pair of Witt algebras. Moreover, together with all the $Q_g$ as appearing in \eqref{Qg} for functions $g$ supported on $S^2$, they form the BMS$_4$ algebra. The generators $L_m, \bar{L}_m$ for $m = 0,  \pm 1$ are the conformal isometry generators on the $S^2$ and all higher and lower modes are named as the super-rotations.
\end{itemize}
The symmetry generators defined for generic cases \eqref{gen} means departure from locality in terms of position space. That is why geometric intuition does not come in handy when trying to find these apparently hidden symmetries even for the trivial case of free field theory.

Despite this very large amount of global symmetry generators being found relatively easily in free theories, for interacting theories (which are described by local Lagrangians), it is an extremely difficult task if not impossible, to find ones beyond those associated with spacetime Killing symmetries and 
internal symmetries. We make this statement even keeping in mind the recently discovered `hidden' Yangian symmetries \cite{Beisert:2017pnr, Beisert:2018zxs} for relativistic supersymmetric theories, which strictly act non-locally on fields.

In contrast, in the present article we have demonstrated an infinite dimensional symmetry algebra (CCA), now acting locally in real space on fields, and thus having well understood geometric interpretation for an interacting theory. To the best of our knowledge only other example of such a symmetry group was in the case of Carrollian scalar electrodynamics in the electric limit \cite{Bagchi:2019clu}.
%
%
\section{Photon propagator in position space}\label{asasasa}
We will begin with the  Lagrangian that contains the gauge fixing term. It is given by
\bea{}\label{lahg} \mathcal{L}=-\frac{1}{4}F^{\mu\nu}F_{\mu\nu}-\frac{1}{2\xi}(\p_{\mu}A^{\mu})^2.\eea
We will now write down the equation (which can be found from \eqref{lahg})
\bea{}\Big(-k^2\eta^{\mu\nu}+(1-\frac{1}{\xi})k^{\mu}k^{\nu}\Big)G_{\nu\rho}(k)=i\delta^{\mu}_{\rho},\eea
from which the expression for the propagator will be obtained. The solution comes out to be
\bea{} G_{\mu\nu}(k)=\frac{-i}{k^2}\Big(\eta_{\mu\nu}-(1-\xi)\frac{k_{\mu}k_{\nu}}{k^2}\Big).\eea
Finally, we can write the most general expression for the photon propagator given by
\bea{}\label{QED}G_{\mu\nu}(x-y):=\langle0|A_{\mu}(x)A_{\nu}(y)|0\rangle=\int^{\infty}_{-\infty} \frac{d^4k}{(2\pi)^4}\, \Big[\frac{-i}{k^2}\Big(\eta_{\mu\nu}-(1-\xi)\frac{k_{\mu}k_{\nu}}{k^2}\Big)\Big]e^{-ik_{\rho}x^{\rho}}. \non\eea
Here, $\mu=(0,1,2,3), k^2=-k_0^2+|\vec{k}|^2$ and $k_{\mu}$ is the four-momentum. For doing the computations, $\xi$ can take any possible value. Some of the popular choices are $\xi=0~\text{(Landau gauge)}; \xi=1~\text{(Feymann gauge)}$.

In this paper, we will only quote the results for the case where $\xi$ is arbitrary while writing the propagator in position space. We will look at both the time-like $(|x^0-y^0|>|\vec{x}-\vec{y}|)$ as well the space-like $(|\vec{x}-\vec{y}|>|x^0-y^0|)$ case. 
The results in \textbf{causal} interval are given as (taking $y=0$ to make calculations bit easy):
\begin{table}[H]
\begin{center}
\begin{tabular}{ |p{.5cm}|p{4cm}|p{5.2cm}|}
 \hline
& Correlators &Results\\
 \hline
 1.  &$ \langle0|A_{0}(x)A_{0}(0)|0\rangle$&$=\Big[-\frac{1}{4\pi^2t^2}+(1-\xi)\frac{1}{8\pi t^2}\Big]$\\
2. &$ \langle0|A_{0}(x)A_{i}(0)|0\rangle$ & $=0$\\
3. & $\langle0|A_{i}(x)A_{j}(0)|0\rangle$  & $=\Big[\delta_{ii}\frac{1}{4\pi^2t^2}+(1-\xi)\frac{1}{8\pi t^2}\Big] ~~\text{for}~~ i=j$ \\
 &&$=0~~\text{for}~~ i\neq j$\\
\hline
\end{tabular}
\caption{Summary of results in causal case}\label{tab:jk4}
\end{center}
\end{table}
Similarly, for the results in \textbf{acausal} case:
\begin{table}[H]
\begin{center}
\begin{tabular}{ |p{.5cm}|p{4cm}| p{5.9cm}|}
 \hline
& Correlators & Results\\
 \hline
1.  &$\langle0|A_{0}(x)A_{0}(0)|0\rangle$&$=\Big[-\frac{1}{4\pi^2r^2}+(1-\xi)\frac{1}{16\pi^2r^2}\Big]$\\
2.&  $\langle0|A_{0}(x)A_{i}(0)|0\rangle$  & $=0$\\
3. &$\langle0|A_{i}(x)A_{j}(0)|0\rangle $ & $ \tiny{=\Big[\frac{1}{4\pi^2r^2}+\frac{(1-\xi)}{16\pi^4}\Big(\frac{r^2-2x_ix_i}{r^4}\Big)\Big]}$\,for $i=j$\\
& & $=(1-\xi)\frac{1}{8\pi^4}\Big(\frac{x_ix_j}{r^4}\Big)~~\text{for}~~ i\neq j$\\
 \hline
\end{tabular}
\caption{Summary of results in acausal case.}\label{tab:title3} 
\end{center}
\end{table}
To get the final results, we have used $x^0=t$ and $r^2=x^i x_i$ in the intermediate steps. Here we have considered the operators to be inserted at same time-slice. Now, one can see that going to the Carrollian limit ie. $t \rightarrow \epsilon \,t, ~ x_i \rightarrow x_i$ does not alter the space-like case above. We use this observation in the Table \ref{tab:jk1} to understand two point correlation functions found from global symmetry arguments.
\bibliographystyle{unsrt} 
\bibliography{ref}

\begin{thebibliography}{10}

\bibitem{Belavin:1984vu}
A.~A. Belavin, Alexander~M. Polyakov, and A.~B. Zamolodchikov.
\newblock {Infinite Conformal Symmetry in Two-Dimensional Quantum Field
  Theory}.
\newblock {\em Nucl. Phys.}, B241:333--380, 1984.
\newblock [,605(1984)].

\bibitem{Simmons-Duffin:2016gjk}
David Simmons-Duffin.
\newblock {The Conformal Bootstrap}.
\newblock In {\em {Proceedings, Theoretical Advanced Study Institute in
  Elementary Particle Physics: New Frontiers in Fields and Strings (TASI 2015):
  Boulder, CO, USA, June 1-26, 2015}}, pages 1--74, 2017.

\bibitem{Duval:2009vt}
Christian Duval and Peter~A. Horvathy.
\newblock {Non-relativistic conformal symmetries and Newton-Cartan structures}.
\newblock {\em J. Phys.}, A42:465206, 2009.

\bibitem{Bagchi:2014ysa}
Arjun Bagchi, Rudranil Basu, and Aditya Mehra.
\newblock {Galilean Conformal Electrodynamics}.
\newblock {\em JHEP}, 11:061, 2014.

\bibitem{Bergshoeff:2014uea}
Eric~A. Bergshoeff, Jelle Hartong, and Jan Rosseel.
\newblock {Torsional Newton Cartan geometry and the Schrodinger algebra}.
\newblock {\em Class. Quant. Grav.}, 32(13):135017, 2015.

\bibitem{Duval:2014uva}
C.~Duval, G.W. Gibbons, and P.A. Horvathy.
\newblock {Conformal Carroll groups and BMS symmetry}.
\newblock {\em Class. Quant. Grav.}, 31:092001, 2014.

\bibitem{Duval:2014uoa}
C.~Duval, G.~W. Gibbons, P.~A. Horvathy, and P.~M. Zhang.
\newblock {Carroll versus Newton and Galilei: two dual non-Einsteinian concepts
  of time}.
\newblock {\em Class. Quant. Grav.}, 31:085016, 2014.

\bibitem{Duval:2014lpa}
C.~Duval, G.~W. Gibbons, and P.~A. Horvathy.
\newblock {Conformal Carroll groups}.
\newblock {\em J. Phys.}, A47(33):335204, 2014.

\bibitem{Bagchi:2015qcw}
Arjun Bagchi, Rudranil Basu, Ashish Kakkar, and Aditya Mehra.
\newblock {Galilean Yang-Mills Theory}.
\newblock {\em JHEP}, 04:051, 2016.

\bibitem{Bergshoeff:2015sic}
Eric Bergshoeff, Jan Rosseel, and Thomas Zojer.
\newblock {Non-relativistic fields from arbitrary contracting backgrounds}.
\newblock {\em Class. Quant. Grav.}, 33(17):175010, 2016.

\bibitem{Hartong:2015xda}
Jelle Hartong.
\newblock {Gauging the Carroll Algebra and Ultra-Relativistic Gravity}.
\newblock {\em JHEP}, 08:069, 2015.

\bibitem{Bergshoeff:2015uaa}
Eric Bergshoeff, Jan Rosseel, and Thomas Zojer.
\newblock {Newton-Cartan (super)gravity as a non-relativistic limit}.
\newblock {\em Class. Quant. Grav.}, 32(20):205003, 2015.

\bibitem{Bagchi:2016bcd}
Arjun Bagchi, Rudranil Basu, Ashish Kakkar, and Aditya Mehra.
\newblock {Flat Holography: Aspects of the dual field theory}.
\newblock {\em JHEP}, 12:147, 2016.

\bibitem{Bergshoeff:2017btm}
Eric Bergshoeff, Joaquim Gomis, Blaise Rollier, Jan Rosseel, and Tonnis ter
  Veldhuis.
\newblock {Carroll versus Galilei Gravity}.
\newblock {\em JHEP}, 03:165, 2017.

\bibitem{Bagchi:2017yvj}
Arjun Bagchi, Joydeep Chakrabortty, and Aditya Mehra.
\newblock {Galilean Field Theories and Conformal Structure}.
\newblock {\em JHEP}, 04:144, 2018.

\bibitem{Basu:2018dub}
Rudranil Basu and Udit~Narayan Chowdhury.
\newblock {Dynamical structure of Carrollian Electrodynamics}.
\newblock {\em JHEP}, 04:111, 2018.

\bibitem{Bagchi:2019clu}
Arjun Bagchi, Rudranil Basu, Aditya Mehra, and Poulami Nandi.
\newblock {Field Theories on Null Manifolds}.
\newblock {\em JHEP}, 02:141, 2020.

\bibitem{Bagchi:2019xfx}
Arjun Bagchi, Aditya Mehra, and Poulami Nandi.
\newblock {Field Theories with Conformal Carrollian Symmetry}.
\newblock {\em JHEP}, 05:108, 2019.

\bibitem{Banerjee:2019axy}
Kinjal Banerjee, Rudranil Basu, and Akhila Mohan.
\newblock {Uniqueness of Galilean Conformal Electrodynamics and its Dynamical
  Structure}.
\newblock {\em JHEP}, 11:041, 2019.

\bibitem{Gupta:2020dtl}
Nishant Gupta and Nemani~V. Suryanarayana.
\newblock {Constructing Carrollian CFTs}.
\newblock 1 2020.

\bibitem{Chapman:2020vtn}
Shira Chapman, Lorenzo Di~Pietro, Kevin~T. Grosvenor, and Ziqi Yan.
\newblock {Renormalization of Galilean Electrodynamics}.
\newblock 7 2020.

\bibitem{Bergshoeff:2014jla}
Eric Bergshoeff, Joaquim Gomis, and Giorgio Longhi.
\newblock {Dynamics of Carroll Particles}.
\newblock {\em Class. Quant. Grav.}, 31(20):205009, 2014.

\bibitem{Bondi:1962px}
H.~Bondi, M.~G.~J. van~der Burg, and A.~W.~K. Metzner.
\newblock {Gravitational waves in general relativity. 7. Waves from
  axisymmetric isolated systems}.
\newblock {\em Proc. Roy. Soc. Lond.}, A269:21--52, 1962.

\bibitem{Sachs:1962zza}
R.~Sachs.
\newblock {Asymptotic symmetries in gravitational theory}.
\newblock {\em Phys. Rev.}, 128:2851--2864, 1962.

\bibitem{Bagchi:2010eg}
Arjun Bagchi.
\newblock {Correspondence between Asymptotically Flat Spacetimes and
  Nonrelativistic Conformal Field Theories}.
\newblock {\em Phys. Rev. Lett.}, 105:171601, 2010.

\bibitem{Campiglia:2015yka}
Miguel Campiglia and Alok Laddha.
\newblock {New symmetries for the Gravitational S-matrix}.
\newblock {\em JHEP}, 04:076, 2015.

\bibitem{Strominger:2017zoo}
Andrew Strominger.
\newblock {Lectures on the Infrared Structure of Gravity and Gauge Theory}.
\newblock 2017.

\bibitem{Laddha:2017ygw}
Alok Laddha and Ashoke Sen.
\newblock {Sub-subleading Soft Graviton Theorem in Generic Theories of Quantum
  Gravity}.
\newblock {\em JHEP}, 10:065, 2017.

\bibitem{Banerjee:2018fgd}
Shamik Banerjee.
\newblock {Symmetries of free massless particles and soft theorems}.
\newblock {\em Gen. Rel. Grav.}, 51(9):128, 2019.

\bibitem{Banerjee:2019tam}
Shamik Banerjee and Pranjal Pandey.
\newblock {Conformal properties of soft-operators. Part II. Use of
  null-states}.
\newblock {\em JHEP}, 02:067, 2020.

\bibitem{Laddha:2019yaj}
Alok Laddha and Ashoke Sen.
\newblock {Classical proof of the classical soft graviton theorem in D>4}.
\newblock {\em Phys. Rev. D}, 101(8):084011, 2020.

\bibitem{Bagchi:2012cy}
Arjun Bagchi and Reza Fareghbal.
\newblock {BMS/GCA Redux: Towards Flatspace Holography from Non-Relativistic
  Symmetries}.
\newblock {\em JHEP}, 10:092, 2012.

\bibitem{Bagchi:2013hja}
Arjun Bagchi and Daniel Grumiller.
\newblock {Holograms of flat space}.
\newblock {\em Int. J. Mod. Phys.}, D22:1342003, 2013.

\bibitem{Bagchi:2012xr}
Arjun Bagchi, St{\'e}phane Detournay, Reza Fareghbal, and Joan Simon.
\newblock {Holography of 3D Flat Cosmological Horizons}.
\newblock {\em Phys. Rev. Lett.}, 110(14):141302, 2013.

\bibitem{Bagchi:2015nca}
Arjun Bagchi, Shankhadeep Chakrabortty, and Pulastya Parekh.
\newblock {Tensionless Strings from Worldsheet Symmetries}.
\newblock {\em JHEP}, 01:158, 2016.

\bibitem{Beisert:2017pnr}
Niklas Beisert, Aleksander Garus, and Matteo Rosso.
\newblock {Yangian Symmetry and Integrability of Planar N=4 Supersymmetric
  Yang-Mills Theory}.
\newblock {\em Phys. Rev. Lett.}, 118(14):141603, 2017.

\bibitem{Beisert:2018zxs}
Niklas Beisert, Aleksander Garus, and Matteo Rosso.
\newblock {Yangian Symmetry for the Action of Planar $\mathcal N=$ 4 Super
  Yang-Mills and $\mathcal N=$ 6 Super Chern-Simons Theories}.
\newblock {\em Phys. Rev.}, D98(4):046006, 2018.

\bibitem{Ciambelli:2018xat}
Luca Ciambelli, Charles Marteau, Anastasios~C. Petkou, P.~Marios Petropoulos,
  and Konstantinos Siampos.
\newblock {Covariant Galilean versus Carrollian hydrodynamics from relativistic
  fluids}.
\newblock {\em Class. Quant. Grav.}, 35(16):165001, 2018.

\bibitem{Levy1965}
Jean-Marc Lévy-Leblond.
\newblock Une nouvelle limite non-relativiste du groupe de poincaré.
\newblock {\em Annales de l'I.H.P. Physique théorique}, 3(1):1--12, 1965.

\bibitem{davis2:1929}
D.~R. Davis.
\newblock {The Inverse Problem of the Calculus of Variations in a Space of
  (n+1) Dimensions}.
\newblock {\em Bull. Amer. Math. Soc.}, {35}{(3)}:371--380, 1929.

\bibitem{jessedouglas:1941}
J.~Douglas.
\newblock {Solution of the Inverse Problem of the Calculus of Variations}.
\newblock {\em Trans. Amer. Math. Soc.}, {50}:71--128, 1941.

\bibitem{Henneaux:1984ke}
M.~Henneaux.
\newblock {ON THE INVERSE PROBLEM OF THE CALCULUS OF VARIATIONS IN FIELD
  THEORY}.
\newblock {\em J. Phys.}, A17:75--85, 1984.

\bibitem{Witten:2010zr}
Edward Witten.
\newblock {A New Look At The Path Integral Of Quantum Mechanics}.
\newblock 9 2010.

\bibitem{Woodhouse:1980pa}
N.~Woodhouse.
\newblock {\em {Geometric Quantization}}.
\newblock 1 1980.

\bibitem{Cao:2018}
Cao Yuan, Fatemi Valla, Demir Ahmet, and et~al.
\newblock {Correlated insulator behaviour at half-filling in magic-angle
  graphene superlattices}.
\newblock {\em Nature}, 556:80–84, 2018.

\end{thebibliography}
\end{document}